\documentclass[pdflatex,sn-nature]{sn-jnl}
\usepackage{graphicx}
\usepackage{placeins}
\usepackage{amsmath,amssymb}
\usepackage{xcolor}
\usepackage{textcomp}
\usepackage{booktabs,array}
\usepackage{xurl}
\usepackage{tabularx}

\usepackage[T1]{fontenc}
\usepackage{newunicodechar}
\newunicodechar{−}{\ensuremath{-}}
\newunicodechar{µ}{\ensuremath{\mu}}
\newunicodechar{ρ}{\ensuremath{\rho}}
\newunicodechar{√}{\ensuremath{\surd}}
\newunicodechar{×}{\ensuremath{\times}}
\newunicodechar{≥}{\ensuremath{\geq}}
\newunicodechar{≤}{\ensuremath{\leq}}
\newunicodechar{–}{--}
\newunicodechar{—}{---}
\newunicodechar{’}{'}
\newunicodechar{α}{\ensuremath{\alpha}}
\newunicodechar{β}{\ensuremath{\beta}}
\newunicodechar{‖}{\ensuremath{\|}}
\newunicodechar{⟨}{\ensuremath{\langle}}
\newunicodechar{⟩}{\ensuremath{\rangle}}
\newunicodechar{Σ}{\ensuremath{\Sigma}}
\newsavebox{\tbox}
\hypersetup{pdftitle={Cell lines that share a well are not independent observations in the Tahoe-100M drug screen},pdfauthor={Xuanrui Huang}}
\usepackage{etoolbox}
\makeatletter
\patchcmd{\@maketitle}{Corresponding author(s). E-mail(s): }{Corresponding author. E-mail: }{\typeout{PATCHOK title}}{\typeout{PATCHFAIL title}}
\patchcmd{\email}{\textcolor{blue}{#1};\ }{\textcolor{blue}{#1}}{\typeout{PATCHOK email}}{\typeout{PATCHFAIL email}}
\makeatother
\begin{document}
\title[Cell lines in one well are not independent in Tahoe-100M]{Cell lines that share a well are not independent observations in the Tahoe-100M drug screen}
\author*{\fnm{Xuanrui} \sur{Huang}}\email{huangxuanrui@u.nus.edu}
\affil{\orgdiv{Department of Biomedical Informatics, Yong Loo Lin School of Medicine}, \orgname{National University of Singapore}, \orgaddress{\country{Singapore}}}
\abstract{Pooled drug screens treat and sequence dozens of cancer cell lines together in one well. Tahoe-100M, the largest, is a main resource for training and benchmarking models that predict how unmeasured cell lines respond to drugs, and its benchmarks, reproducibility ceilings and statistical tests treat the cell lines of a well, and random halves of a well's cells, as independent observations of a drug's effect. Because the drug is applied to the well, this is harmless only if the lines of a well share nothing beyond the drug's reproducible effect. We tested this with pairs of untreated control wells on 13 plates, with replicate wells whose labels we checked against their partners, and with two plates that repeat each other well by well. In each test the cell lines of a well share a component of expression that a replicate well does not reproduce, also when the two wells occupy the same inferred position; where drugs are present, this component cannot be told apart from an interaction of plate and drug, an ambiguity from which the untreated control wells are free. In replicate wells at different positions whose labels pass the check, about a third of what two halves of a well have in common (35\%, 95\% interval 22--58\%, 9 compounds) is not reproduced by the replicate well. The well effect inflates results that take something from a well and judge it in the same well: retrained versions of two published models that take a measured response as input lose about a quarter to a third of their score when the input comes from a replicate well instead of the predicted well (22--43\% under the control-well reference), halves of a well overstate how well the experiment reproduces, and a test across the lines of a well calls 5 of 13 plates significant when two untreated wells are compared. Among strongly responding replicate wells, many labels, those of plate 13 above all, fail a check against their partners. Evaluations, ceilings and tests on such data should treat the well as the experimental unit.}
\maketitle

\section{Introduction}

Predicting how cells respond to drugs is a central task in drug discovery and in recent work on ``virtual cell'' models: a model that infers the response of an unmeasured cell line from measured ones could replace many experiments\cite{bunne2024,roohani2025}. Such models need genome-wide transcriptional responses to many drugs in many cell lines, and pooled drug screens provide them. They culture, treat and sequence dozens of cell lines together in one well and assign each sequenced cell to its line afterwards from its genotype, so one well yields the responses of dozens of lines\cite{mcfarland2020,zhang2026}. The largest such screen, Tahoe-100M, treated about 50 cancer cell lines with about 380 drugs at three doses on 14 plates of 96 wells and profiled about 100 million cells\cite{zhang2026}. It is a main resource for training and evaluating these models and has been used for conclusions about drug action\cite{zhang2026,shen2026}.

What such data can show depends on what counts as an independent observation. In a designed experiment the experimental unit is the smallest unit to which a treatment is independently applied; observations within one unit are subsamples, and treating them as independent replicates, called pseudoreplication, makes estimates look more precise and agreement look stronger than they are\cite{hurlbert1984}. In single-cell studies, for example, the cells of one donor are subsamples of that donor, and tests that treat them as independent inflate false discoveries\cite{zimmerman2021,squair2021}. In a pooled drug screen the drug is applied to a well, so the well is the experimental unit and the cell lines in it are subplots, as in a split-plot design. The quantity that models and benchmarks aim at is a drug's effect on a cell line, which another experiment would reproduce; anything particular to one well is not part of it. Treating the lines of a well as independent observations of that effect is harmless if they share nothing beyond it. Whether they do is an empirical question.

Several practices built on Tahoe-100M assume that the lines of a well share nothing beyond the drug. A widely used evaluation, introduced with the State model, holds out five cell lines while the responses of the other lines to the same drugs, including those measured in the same wells as the held-out lines, enter training\cite{adduri2025}; models from the group that generated the data use the same split\cite{gandhi2025,rhaister2026}. Other models represent a drug by its measured response in another cell line, which usually comes from the well being predicted\cite{kaufmann2026,molina2026}. The agreement between two random halves of a well's cells, corrected for the halved cell number, is used as an upper bound on any model's score\cite{celleval}, and the same halves are used to decide which measurements are reliable\cite{wang2026}. Biological analyses treat the lines of a well as independent samples, for example when testing whether a response is shared across lines\cite{zhang2026,shen2026}.

Pooling was designed to give all lines in a well the same handling, so that differences between them reflect the lines themselves\cite{mcfarland2020}. The same design makes the premise doubtful. A well is a separate culture: its cells are seeded, treated, cultured, dissociated, fixed and marked with a first-round barcode that identifies the well before they are pooled with the cells of other wells for sequencing. Each of these steps can deviate from one well to the next, and all lines in a well go through the same deviations, so they may change together. In imaging and other high-throughput screens, measurements from the same well position resemble each other more than measurements from different positions\cite{mazoure2017,kalinin2025}. We call a component of a well's expression beyond the drug's reproducible effect that a replicate well at another position does not reproduce a well effect. A replicate at the same position may reproduce part of it, and part of it may be shared by the lines of the well. If a well effect exists, information leaks between training and test data whenever both come from one well: two random halves of a well both contain the effect and appear to agree, the other lines of a well contain it and appear to predict a held-out line, and a response shared across lines may partly be the well's.

Here we test whether the cell lines of a well share a well effect in Tahoe-100M, and what it does to the practices above. The test asks whether the other lines of a well predict a line's response better than the corresponding lines of another well. We apply it first where no drug acts, comparing the two untreated control wells of each plate with those of other plates, and then to replicate wells of the same drug and dose after verifying their labels; two of the plates repeat each other well by well, with the same drug, dose and position. We then examine explanations other than a well effect, estimate how much of a measurement it accounts for, and measure the inflation of model scores, reproducibility ceilings and statistical tests. Finally we compare two screens of other designs.

\section{Results}

Throughout, a cell line's response in a well is its mean expression there minus its mean in a reference, and the response is predicted by the mean response of the other cell lines in the same well, the perturbation-mean baseline of the State study\cite{adduri2025}, which uses no data of the predicted line and needs no training. Accuracy is the Pearson correlation between prediction and response over the genes expressed in the line. Published analyses take the plate's control wells as the reference; because the control wells have their own well effect, which a line and the lines predicting it would then share, we use the line's mean over the plate's other treated wells instead (treated-well reference), so that effects common to a plate, or to a line on a plate, cancel, and we compare the control-well reference where it matters (Methods). Plate 13 is excluded throughout: its replicate labels mostly fail the check described below, and its control wells sit at other positions than on every other plate.

Where a drug--dose condition was measured in two wells, the same response can be predicted twice: from the other lines in the predicted line's own well, and from the other lines in the replicate well. If the lines of a well share nothing beyond the drug, and the two wells received the drug at the same strength, the two predictions are equally accurate on average. The advantage of the same-well prediction therefore measures what the lines of one well share and its replicate does not, whatever the source; differences in drug strength between the wells are examined below.

\subsection{Cell lines in untreated wells share a well effect}

\begin{figure}[tbp]
\centering
\makebox[\textwidth][c]{\includegraphics{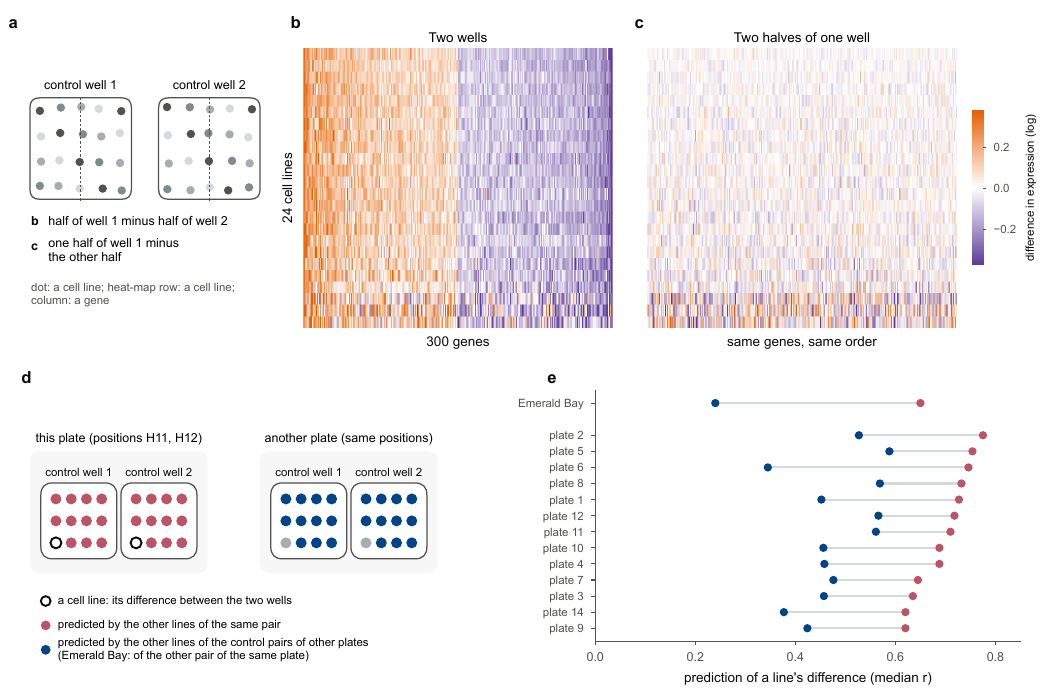}}
\caption{\textbf{The cell lines of an untreated well shift together; the two halves of one well do not differ that way.} \textbf{a}, The two comparisons of \textbf{b} and \textbf{c}, on the two untreated control wells (DMSO) of one plate. \textbf{b}, Each row is one cell line's difference in expression between half of its cells in control well 1 and half of its cells in control well 2; each column is a gene. \textbf{c}, The same lines and genes: difference between the two halves of the line's cells in control well 1. Halves are random halves of each line's cells; profiles are $\log(1+10^4c/N)$ of counts summed over the cells ($c$, a gene's count; $N$, the total). The 300 genes are the 150 with the highest and the 150 with the lowest mean whole-well difference over 24 randomly chosen other lines, ordered by that mean; the 24 lines shown were not used for the choice and are ordered by the noise of their difference between the two halves of one well (quietest on top). Plate 1, the plate at the median of the 13 by the median correlation of a shown line with the choosing lines' mean over these genes, which is 0.89 in \textbf{b} and 0.00 in \textbf{c} (all plates: Supplementary Fig.~\ref{figS1}). \textbf{d}, The prediction: a line's difference between the two control wells of a plate is predicted by the mean difference of the other lines of the same pair (red) or of the other lines of the control pairs of the other plates, at the same two positions (blue); the target line itself (grey) is left out. \textbf{e}, Median correlation over the lines of each plate, over all expressed genes (plate 13 not used); for Emerald Bay, a later plate of the same platform, the other lines of a pair (red) and the lines of the other pair of the same plate (blue), over its three splits of four control wells into two pairs (Supplementary Table~\ref{tabCtrl}).}\label{fig1}
\end{figure}

The cleanest test needs no drug. Every plate has two control wells that received only the solvent, dimethyl sulfoxide (DMSO), and the difference between a line's expression in the two contains no drug effect. If the lines of a well were independent, these differences would be unrelated between lines. They are not: across genes they rise and fall together in every line, whereas the differences between two random halves of one control well do not (Fig.~\ref{fig1}a--c). The genes shown were chosen and ordered from a random half of the lines and only the other half is shown, so the pattern cannot come from the choice of genes; and both comparisons set half a well against half a well, so their sampling noise is the same. Over these genes, the median correlation of a shown line's difference with the mean difference of the other half of the lines is 0.82 to 0.94 on the 13 plates, against $-0.03$ to 0.04 between halves of one well (Supplementary Fig.~\ref{figS1}); the genes were chosen as those that differ most, so these values show the contrast with halves rather than the size of the effect, and thinning the deeper well to the depth of the other changes it by at most 0.01.

Over all expressed genes, the other lines of a pair predict a line's difference at a median correlation of 0.70 (48 lines, 13 plates; Fig.~\ref{fig1}d,e). The control pairs of the other plates also predict it, at 0.47: part of the shared difference belongs to the wells' position, or to something tied to position such as the first-round barcode, because on these plates the two control wells sit at the same two positions, H11 and H12, as inferred from sample numbers (Methods). The rest belongs to the plate's own two wells: on each of the 13 plates the pair's own lines predict better than the pairs of the other plates at the same positions. The excess holds in a permutation test with the plate as the unit, and with counts summed over cells or at matched depth (Supplementary Table~\ref{tabCtrl}). The sharing itself holds on data we had not used, the four DMSO wells of Emerald Bay, a later plate of the same platform\cite{emeraldbay}: the other lines of a pair predict a line's difference at 0.65, and the lines of the other pair of the same plate at 0.24 (Supplementary Table~\ref{tabCtrl}); the positions of these wells were not checked, so this plate cannot separate the part tied to position. Without any drug, the lines of a well thus share a component beyond the drug: part of it is tied to position and recurs in wells at the same position on other plates, and part of it is a well effect that such wells do not reproduce.

\subsection{Drug-treated replicate wells show the same pattern}

\begin{figure}[tbp]
\centering
\makebox[\textwidth][c]{\includegraphics{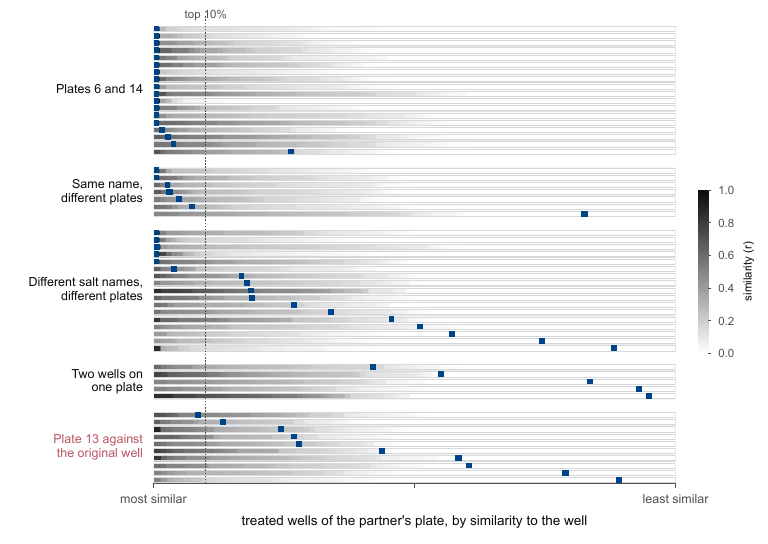}}
\caption{\textbf{Replicate labels need checking, and those of plate 13 mostly fail.} Each row is one pair of replicate wells in which at least one well responds strongly (response length $\geq$ 3). Grey cells: similarity (Pearson correlation of the line-averaged responses) of one well of the pair to every treated well of the other well's plate, sorted from most to least similar; blue: the labelled replicate, placed at its rank. Of the two directions, the one in which the replicate ranks lower is shown, so a blue cell left of the dotted line (top 10\%) means that the two wells match. Rows are sorted by that rank. Plates 6 and 14 (18 pairs; their wells also share a position, which the rank detects as well), same-name pairs between plates 1--3 and 7--12 at different positions (7), pairs with different salt names on different plates (17), two wells on one plate (5) and plate 13 against the well on another plate (10); counts in Supplementary Table~\ref{tabS1}.}\label{fig2}
\end{figure}

Testing for a well effect where drugs act requires replicate wells, and such a test is only as good as the replicate labels. Tahoe-100M contains several kinds of replicates: plate 14 repeats plate 6 with the same drug, dose and position in each well (positions are inferred from sample numbers; Methods); 18 compounds appear under different salt names on different plates and positions; three drugs of plates 1--3 are repeated at other positions on plates 7--12; a few drugs occupy two wells of one plate; and plate 13 repeats 45 drug--dose conditions of other plates at other positions. Where the response is strong, a label can be checked, because two wells with a strong response to the same drug should resemble each other more than they resemble the plate's other wells. We call two wells matched if each ranks the other among the 10\% most similar treated wells of the other's plate (Methods). Among pairs in which a well responds strongly, meaning that the norm of its line-averaged response is at least 3 (Methods), the two wells match in 17 of 18 pairs on plates 6 and 14 and in 6 of 7 same-name pairs at different positions, but in only 6 of 17 pairs with different salt names (10 compounds), in none of 5 pairs of two wells on one plate (4 compounds) and in 1 of 10 pairs between plate 13 and other plates (7 compounds; Fig.~\ref{fig2}; Supplementary Table~\ref{tabS1}). A failed check does not prove a wrong label, because a strong response can also come from the well alone, but these labels cannot be taken on trust. The wells of plate 13 also respond weakly: where the first well of a pair responds strongly, its plate-13 partner reaches a median of 0.44 of its response (6 pairs, 5 compounds), against 0.71 to 0.97 for the other kinds of replicates. Plate 13 received drug, since adagrasib, present on every plate, acts there in sensitive lines; whether its records, dosing or handling are at fault, the public data cannot tell.

Most pairs respond too weakly for this check, so we checked every pair with two readouts that do not depend on each other: the line-averaged expression response and a viability readout, the change in each line's share of the well's cells. A pair is concordant if its wells match by either readout. Pairs of wells of two different compounds pass this check in 10--12\% of cases (25\% for adjacent wells on one plate), so concordance indicates correct labels only where many more pairs pass than this. At different positions, 18 of 65 pairs (9 compounds) are concordant, against 6.4 expected if every label were wrong; on plates 6 and 14, 80 of 90; for two wells on one plate, 6 of 24, which cannot be told apart from chance (expected 3.0; $p=0.07$, treating the pairs as independent; Supplementary Table~\ref{tabS1}).

\begin{figure}[tbp]
\centering
\makebox[\textwidth][c]{\includegraphics{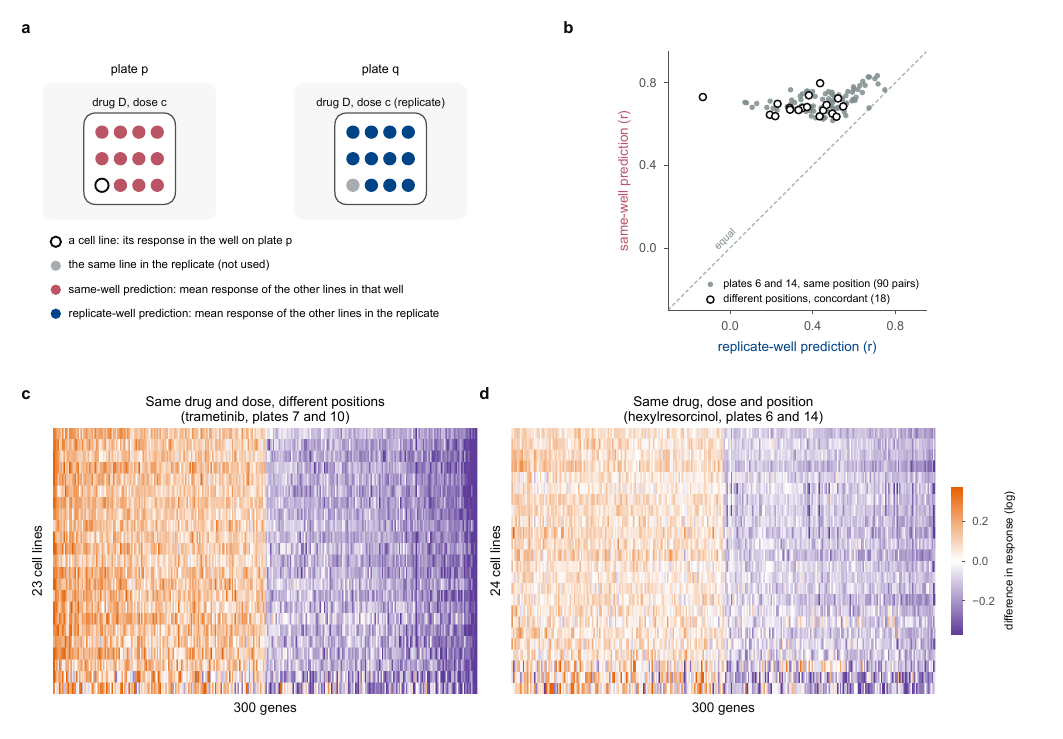}}
\caption{\textbf{In drug-treated replicate wells the other lines of the same well predict a line better than those of the replicate well.} \textbf{a}, The comparison: a line's response in one well is predicted by the mean response of the other lines in the same well (red) or in the replicate well, which received the same drug at the same dose (blue); the target line's own response in the replicate (grey) is not used. \textbf{b}, Same-well against replicate-well prediction for each pair of wells (median over its lines, both wells as the target's well): the 90 pairs of plates 6 and 14, at the same position, and the 18 concordant pairs at different positions. Every pair lies above the line of equality. \textbf{c},~\textbf{d}, As Fig.~\ref{fig1}b: each row is one cell line's difference in response between half of its cells in one replicate well and half of its cells in the other, each against its own plate's other treated wells; 300 genes chosen and ordered from 24 other lines; lines ordered by the noise of their difference between halves of one well; shown are the lines of the second set measured in both wells (23 in \textbf{c}, 24 in \textbf{d}). \textbf{c}, Trametinib at 0.05\,µM on plates 7 and 10, at different positions: among the 9 concordant pairs whose lines' responses agree between the two wells at least as well as the median of the 18 (0.57; this pair 0.79), so that the drug response reproduces, the pair at the median of the correlation given below. \textbf{d}, Hexylresorcinol at 5\,µM on plates 6 and 14, at the same position: the pair at the median of the 90. Median correlation of a shown line with the choosing lines' mean over these genes: 0.92 in \textbf{c} and 0.80 in \textbf{d} ($-0.01$ and $-0.02$ between halves of one well); over the 18 concordant pairs at different positions 0.83 to 0.95, and over the 90 pairs of plates 6 and 14 0.62 to 0.90.}\label{fig3}
\end{figure}

In concordant pairs at different positions, the same-well prediction reaches a median correlation of 0.68 and the replicate-well prediction 0.38, and the same well predicts better in each of the 18 pairs and for each of the 9 compounds (Fig.~\ref{fig3}a,b; Supplementary Table~\ref{tabS2}). The difference between the two wells looks like that between two control wells: the lines rise and fall together across genes (Fig.~\ref{fig3}c). The advantage, the mean over lines and pairs of the difference between the two correlations, is 0.32, and 0.38 with each compound weighted equally, which limits the weight of trametinib, the compound with the most pairs (8 of 18). The 95\% interval of the advantage over compounds and plates, 0.18--0.51, rests on only 9 compounds. Swapping each line's responses between the two wells of a pair at random, 1,000 times, never gives more than 0.02, so the two wells differ; why they differ is the subject of what follows. The advantage is 0.29--0.30 when concordance is judged by viability alone, which does not use the scored responses, or from one half of the cells with the scores taken from the other. Its size depends on which pairs pass: looser checks give larger values and a stricter one a smaller value (Supplementary Table~\ref{tabS6}), as expected if looser checks admit wrong labels, which remove the shared drug effect from the replicate well; stricter checks also keep pairs whose wells agree more. Even if the 6.4 pairs expected to pass by chance were all mislabelled and each carried an advantage of 0.7, more than any group shows, the remaining pairs would keep an advantage of about 0.11. At different positions the advantage is therefore positive, about 0.3, but imprecisely sized. Concordant pairs on one plate give 0.69 against 0.17 and are only described.

Two further groups avoid the weaknesses of the labels. Adagrasib at 0.05\,µM was placed on every plate, under one name, and 60 of the 91 pairs of its wells are concordant. For each of its 14 wells the same well predicts better than its other wells (mean advantage 0.37). Plates 6 and 14 hold the same drug, dose and position in each well, and the Tahoe-100M study treats them as replicates\cite{zhang2026}. Over all 90 pairs at the same position, the same-well prediction reaches 0.70 and the replicate-well prediction 0.48 (Supplementary Table~\ref{tabS2}); the same well predicts better in each of the 90 pairs (Fig.~\ref{fig3}b), the lines again share the difference between the two wells (Fig.~\ref{fig3}d), and the advantage is 0.24 (95\% interval 0.22--0.27, from resampling compounds; with one pair of plates, variation between plates is not covered), positive for each of the 87 compounds and 0.19 to 0.22 over the concordant pairs, however concordance is judged. Drug and position are the same in the two wells, so neither can produce this gap; a difference in dosing between the plates is examined below. The difference between the two plates at one position correlates weakly with that at an adjacent position (median 0.15, against $-0.05$ four or more positions away; $p=0.001$), so a small part of the effect is shared by neighbouring wells.

Where the drug has no detectable viability effect in either well, that is, where the spread of the lines' share changes lies within the range seen between two control wells, the gap is largest, but most of these labels cannot be checked: at different positions the same-well prediction reaches 0.73 and the replicate-well prediction 0.08 (24 pairs; Supplementary Table~\ref{tabS2}). The low replicate-well prediction may reflect a weak drug, a wrong label or both; these pairs show only that the same-well prediction stays high where the replicate well holds almost no predictable signal.

\subsection{Alternatives to a well effect}

Three other sources could give the same well an advantage: the reference, a difference in the drug between the two wells, and technical sharing between the lines of one well.

The reference matters in one case. The treated-well reference averages about 90 wells and shares little between a line and the lines that predict it; dividing it into two halves, one subtracted from the target line and one from the other lines, changes the advantage by at most 0.03, and restricting it to treated wells without detectable viability effect, which leaves out most of the plate's average drug effect that a target and its same-well prediction would otherwise share, also changes it by at most 0.03. Under the control-well reference, the target and the same-well prediction subtract the same control wells, whose own well effect favours the same well. Subtracting one control well from every line keeps this shared term, and subtracting one control well from the target and the other from the remaining lines removes it; both versions add a single control well's own effect to the responses, which lowers all correlations and compresses the advantage. At different positions the advantage changes little (0.30 to 0.26), but on the concordant pairs of plates 6 and 14 it falls from 0.23 to 0.08, still positive for 70 of 77 compounds (Supplementary Table~\ref{tabS2}). The shared term weighs most where responses are small, as on plates 6 and 14, where 65 of the 90 pairs have no detectable viability effect; at different positions the pairs without such an effect also lose more (0.45 to 0.37). Under the control-well reference the advantage on the concordant pairs of plates 6 and 14 thus lies between 0.08 and 0.23, depending on how the control wells are assigned, and the difference between the two, 0.15, comes from the shared control well; on the same pairs, the references that share no control well give 0.19 to 0.22.

A drug could act more strongly in one well than in its replicate, from dosing or batch, and the lines of the stronger well would then move together along their own responses. Removing each line's own response direction from its whole-well difference between the two wells leaves the shared pattern in place (median correlation with the mean of the lines that chose the genes, over whole-well differences and therefore higher than the half-well values of Fig.~\ref{fig3}, 0.93 before and 0.92 after in concordant pairs at different positions, 0.85 and 0.84 on plates 6 and 14). Removing the direction in which each line's response changes with dose, estimated from the compound's other two doses, leaves the advantage at 0.21 in the 79 concordant pairs of plates 6 and 14 whose compound has two other doses; this test is informative mainly for strongly responding drugs, whose dose direction is estimated well. Pairs with the same name and pairs with different salt names, which come from different products, have the same advantage (0.32 each), although only 2 compounds have same-name pairs (Supplementary Table~\ref{tabS7}). And the two adagrasib wells of plate 7 agree less than any pair of its wells on different plates, the opposite of what a dosing difference between plates would produce, although this is a single case. Where a drug is present, a well effect cannot in general be told from an interaction of plate and drug, such as a change in the shape of the response with dose, because each plate holds one well of nearly every drug; the control-well pairs, which contain no drug, are free of this ambiguity.

Technical sharing accounts for little of the effect, but contamination is not excluded. Sequencing depth leaves the control-pair result in place, and on plates 6 and 14 regression on quality measures of the line and the well and removal of row and column effects change the advantage by at most 0.01, tested only under the control-well reference. Misassigned cells and ambient RNA would put other lines' responses into a line's measurement and act like a same-well prediction. An estimate from genes that a line barely expresses puts this contamination at about 6\%; it is coarse, because only 3 of the 48 lines have enough such genes and the other lines take the median of their well. Removing it to first order lowers the advantage by at most 0.03 in any group, but in concordant pairs at different positions records with a higher estimate have a larger advantage (Spearman correlation 0.43; 0.40 against 0.25 above and below the median estimate), which the removal does not account for, whereas on plates 6 and 14 and in the control-well pairs no such relation is seen (0.02 and $-0.29$; Supplementary Table~\ref{tabS7}). Because 45 of the 48 lines take the median of their well, this relation rests on few independent comparisons.

\subsection{Where a drug acts, about a third of what two halves of a well share is not reproduced by a replicate well}

How much of a measurement does the well effect account for? Two random halves of a line's cells in one well share the drug effect and the well effect; the line's responses in two replicate wells share only the drug effect. The mean inner product over genes between halves of one well therefore estimates the drug and well components together, and that between halves of two replicate wells the drug component alone. Sampling noise, independent between halves, enters neither, so the well's share is not cell sampling noise and more cells would not remove it (Methods). These estimates correspond to moment estimates of the variance components of a random-effects model with line-by-drug, well and line-by-well terms. They count with the well anything that the replicate well does not reproduce, including an interaction of plate and drug, and a part tied to position is counted with the drug when the replicates share a position.

In concordant pairs at different positions, the well accounts for 35\% of what two halves share (95\% interval 22--58\%), with each line and pair weighted by the size of its response; with each compound weighted equally, for 51\%, a less stable value without an interval, because weak responses enter it as ratios of small numbers (Supplementary Table~\ref{tabS3}); with a reference that leaves out most of the plate's average drug effect, the pooled share is 0.38 (Methods). Mislabelled pairs among those that pass the check would raise these shares. The part shared by the lines of the well is 22\% of the total, and the rest is specific to a line in that well. A proportional difference in drug strength between the two wells would also be counted with the well, but an estimate that such a difference cannot affect is 32\% (Methods). On plates 6 and 14 the share is 30\% (18--49\%); the shared position lowers it, and a dosing difference between the plates would raise it. Where the drug has no detectable viability effect the share is 77--79\%, but there what two halves share is small, the intervals reach beyond 100\% and most labels are unchecked, so these values are only described.

\subsection{Results that take from a well and judge in the same well are inflated}

A well effect does not bias every use of the data. It inflates a result where something is taken from a well and the result is judged in the same well, because the effect is then on both sides of the comparison. We measured this for the four practices of the Introduction.

\begin{figure}[tbp]
\centering
\makebox[\textwidth][c]{\includegraphics{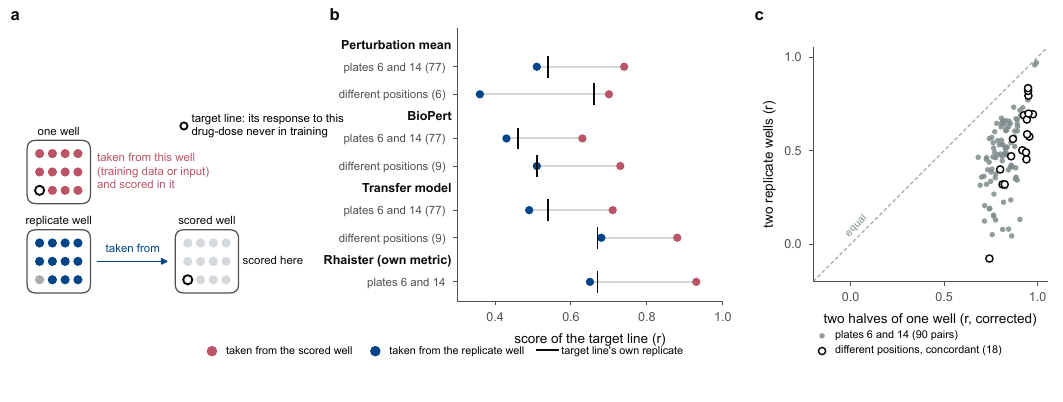}}
\caption{\textbf{Taken from a well and judged in the same well, model scores and reproducibility ceilings are inflated.} \textbf{a}, The comparison: training data or a model's input are taken from the other lines of one well, and the target line, whose response to this drug--dose never entered training, is scored in the same well (red) or in its replicate (blue). \textbf{b}, Concordant pairs, treated-well reference, except for Rhaister; numbers of compounds in parentheses. Perturbation mean: State's five held-out lines, few-shot, with the other lines of one well of each replicated drug--dose in training, scored in that well or in its replicate (median over records; the additive model gives nearly the same values). BioPert and the Molina--Zhang transfer model, with every compound with a replicate well held out: input taken from the target's well or from the replicate well, scored in the target's well. Rhaister: few-shot, on its own metric, the correlation of predicted and observed differential expression with each plate against its own control wells, over all 1,588 records of the drugs repeated on plates 6 and 14, scored against the plate whose wells entered training or against the other plate. Black ticks: the target line's own response in the replicate well (for the perturbation mean, the test lines' agreement between the two wells over the same records; for Rhaister, the agreement of the two plates' observations on its metric). Values and other settings in Supplementary Table~\ref{tabS4}. \textbf{c}, Agreement of a line's responses between two random halves of one well (Spearman--Brown corrected; mean of the two wells) against agreement between the two replicate wells, for each pair (median over its lines). Every pair lies below the line of equality.}\label{fig4}
\end{figure}

\textbf{Training on the other lines of the predicted well.} In the State evaluation, the other lines' responses in the held-out lines' wells enter training. We repeated the evaluation but let only one well of each replicated drug--dose enter training, and scored the held-out lines both against that well, as usual, and against the replicate well, whose lines never entered training (Methods). In this setting the perturbation mean is the same-well prediction of the previous sections, so it restates the advantage in the terms of the benchmark. In the few-shot setting, in which the held-out lines contribute 30\% of their other conditions to training, it scores 0.74 against the training well on the concordant pairs of plates 6 and 14, as high as the ceiling for a prediction that uses nothing from the scored well ($\sqrt{0.54}\approx0.73$, from these lines' own agreement of 0.54 between the two wells; Methods), and 0.51 against the replicate well (77 compounds; Fig.~\ref{fig4}a,b; Supplementary Tables~\ref{tabS4} and \ref{tabCeil}); on the concordant pairs at different positions, about 0.70 and 0.36 (6 compounds). An additive model of line and drug--dose effects gives nearly the same values.

Rhaister, a published model trained with the same split, reports 0.87 on its own metric, the correlation of predicted and observed differential expression\cite{rhaister2026}. Run with its own code and inputs in the few-shot setting on the drugs repeated on plates 6 and 14, it scores 0.93 against the plate whose wells entered training and 0.65 against the other plate (Fig.~\ref{fig4}b; Supplementary Table~\ref{tabS4}). In this setting the target line's own other conditions on the training plate also enter training and share that plate's control wells with the scored response; in the zero-shot setting, in which none of them does, it scores 0.65 and 0.44, still about a third lower against the other plate. The two plates' observed differential expression of the same records correlates at 0.67 on this metric, so a prediction that uses nothing from the scored plate has a ceiling of about $\sqrt{0.67}\approx0.82$ if the two plates are equally noisy (Methods), and less if part of the 0.67 comes from the plates' shared positions. In the few-shot setting, against the training plate, Rhaister exceeds this ceiling. The excess comes from the training plate: from its wells, or from its control wells, which every condition on that plate shares, including the target line's own training conditions in the few-shot setting.

\textbf{Inputs taken from the predicted well.} BioPert represents a drug by the line A549's response to it\cite{kaufmann2026}, and the Molina--Zhang transfer model predicts a line from the mean response of other lines\cite{molina2026}; in Tahoe-100M both inputs usually come from the well being predicted. We held out every compound with a replicate well, retrained or refitted both models, and took each input either from the target's well or from the replicate well (Methods). With the input from the replicate well, BioPert falls from 0.63 to 0.43 on the concordant pairs of plates 6 and 14 and from 0.73 to 0.51 on the concordant pairs at different positions, and the transfer model from 0.71 to 0.49 and from 0.88 to 0.68 (Fig.~\ref{fig4}b; Supplementary Table~\ref{tabS4}). Each model then does about as well as the target line's own response in the replicate well. In pairs without detectable viability effect, whose labels are mostly unchecked, both fall from 0.6--0.7 to about 0.

\textbf{Ceilings from halves of a well.} Two random halves of a well agree at 0.80 to 0.94 after Spearman--Brown correction\cite{spearman1910,brown1910}, which estimates the agreement of whole wells. A line's responses in two whole replicate wells correlate at about 0.5 on plates 6 and 14, where most pairs are concordant and position is shared, and about 0.6 in concordant pairs at different positions, and in each of these pairs below the agreement of two halves (Fig.~\ref{fig4}c; Supplementary Table~\ref{tabCeil}). Halves agree more than two wells do because they also share the well, so a ceiling set by halves overstates how well the experiment reproduces, and part of the gap between model scores and that ceiling cannot be closed by better models.

\textbf{Tests across the lines of a well.} The Tahoe-100M study tested whether cell-cycle inhibitors shift the cell-cycle composition consistently across lines, with a sign-flip test over the lines of a treated well against the plate's control wells\cite{zhang2026}. Between the two untreated control wells of a plate, where no drug acts, the same test calls 5 of 13 plates significant at $p<0.05$, with signs in both directions (Supplementary Table~\ref{tabCeil}). In the original design, applied to 599 wells of 238 compounds (salt names removed) selected as unlikely to move the cell cycle (no detectable viability effect, and no term for the cell cycle, DNA, mitosis, transcription, chromatin or major signalling pathways in their annotated targets and mechanisms, a keyword filter that keeps some kinase and nuclear-receptor drugs), it calls 68\% of the wells significant, an upper bound on its false-positive rate, between 48\% and 93\% on each plate. Some of these drugs may act on the cell cycle a little, and the wells of a plate share its control wells, so these calls are not independent: on some plates most calls go in one direction, as an effect of the shared control wells would produce, and on others both directions are equally common, as the treated wells' own effects would produce. The test detects real differences between wells but attributes them to the drug: for the well's component, the dozens of lines in a well are one observation, not dozens.

\subsection{Screens of other designs}

The OP3 screen pools the six immune cell types of a donor in one well\cite{szalata2024}. In wells with solvent only, the mean of the other cell types predicts a cell type at 0.27 from the same well and 0.00 from another well; for the strong histone deacetylase inhibitor belinostat, at 0.53 and 0.50 (Supplementary Table~\ref{tabS5}). This is consistent with a well effect that is most visible where the drug does little, but it is descriptive, without an interval or a test with the well as the unit, and with five other cell types per well instead of about 47 other lines its size is not comparable with Tahoe-100M; responses are taken against the plate's other solvent wells, leaving the well itself out, which can also lower the prediction from another well. sci-Plex3 places one cell line per well\cite{srivatsan2020}, so there is no borrowing between lines, but two halves of a well may still share the well: at equal cell numbers, two halves of one well agree more than halves of two replicate wells (by 0.031, interval 0.028--0.033; in 84\% of pairs), which a difference in drug strength between replicate wells, or any difference between their plates, would also produce. The design thus decides which practices a well effect can inflate.

\section{Discussion}

In Tahoe-100M the cell lines of a well share a component of expression that a replicate well does not reproduce. The cleanest evidence needs no drug: on each of 13 plates the lines of a control-well pair predict each other's differences better than the control-well pairs at the same positions on other plates, and on a later plate of the same platform each control-well pair predicts its own lines better than the plate's other pair. With drugs, the same holds in concordant replicate wells at different positions, for adagrasib on every plate, and on two plates that repeat each other with the same drug, dose and inferred position, although there it cannot be told apart from an interaction of plate and drug. In concordant replicates at different positions, about a third of what two halves of a well share is not reproduced by the replicate well. The well is the experimental unit by design, and these results show that ignoring this is not harmless: for the well's component, the dozens of lines in a well are one observation.

This is a form of pseudoreplication. In single-cell studies, the cells of one donor share that donor\cite{zimmerman2021,squair2021}; in pooled CRISPR screens, cells carrying one guide RNA descend from a few transduced cells, and random halves share clones\cite{huang2026clone}. Here the subsamples are different cell lines rather than copies of one sample, which makes the problem easy to miss: the lines differ biologically, and predicting one from the others looks like prediction across independent samples. Where the others were measured in the same well, part of what they predict is the well.

The effect inflates a result only where something is taken from a well and the result is judged in the same well: models trained on, or given inputs from, the other lines of the predicted well and scored in it; ceilings set from halves of one well; and tests that treat the lines of a well as independent samples. To fill in an unmeasured line in a well that was measured, borrowing from that well is legitimate; to predict a drug's reproducible effect on a line, as in the benchmarks, it is not. The remedies follow from taking the well as the experimental unit. Models should be scored against wells that did not supply their training data or inputs. Most drug--dose conditions in Tahoe-100M were measured in one well, and among the replicated ones mainly plates 6 and 14 have labels that pass the check, so such scores are available for few conditions; for the others, a condition whose well supplies training data cannot be scored without this leakage, and excluding the well from training turns the task into prediction for an unseen condition. Reproducibility and ceilings should be estimated from separate wells: for a prediction scored against one well that uses nothing from it or from wells that share its well effect, the ceiling is about the square root of the correlation between two replicate wells, assuming equally noisy wells (Methods); replicates at the same position overstate it, and a prediction that borrows from the scored well can exceed it. Statistical tests should use the well as the unit, for instance as a random effect where wells are replicated, with a null distribution taken from wells, not from the lines within one well; between two untreated wells, the test above calls 5 of 13 plates significant. Replicate wells should preferably lie far apart on the plate, because wells at the same or adjacent positions share part of the effect.

How much this matters depends on the condition. Where the drug has little effect, the well's share of a measurement is largest, although most of these labels could not be checked; where the drug acts, the replicate well also carries predictable signal, but the same well still predicts better. Pooling removes batch differences between cell lines at the price of a shared well, and it is because Tahoe-100M contains replicate wells that this can be tested at all. We read the well effect as a consequence of pooling rather than a defect of these data; the replicate labels, those of plate 13 above all, are a separate matter for the data's producers.

The study has limits. The evidence comes from Tahoe-100M and one later plate of the same platform, with two smaller screens of other designs. Concordant replicates at different positions cover 9 compounds, the check also passes 10--12\% of pairs of different compounds, and for two wells on one plate it cannot be told apart from chance. Part of the well effect is tied to position or to the first-round barcode, which the public data cannot separate, and whether it forms before or after the cells of different wells are pooled is unknown. Where a drug is present, an interaction of plate and drug, such as a change in the shape of the response with dose, cannot be separated from the well effect; on plates 6 and 14 in particular, the well effect cannot be fully separated from a difference in dosing between the plates, although the consequence for evaluation is the same. Responses in the main analyses are differences of mean log-normalized expression, and other transforms, such as Pearson residuals, were not tried; on the control-well pairs, log counts per million of summed counts, also at matched depth, give nearly the same advantage (0.19 and 0.18 against 0.20, 13 of 13 plates). Contamination between lines was estimated only coarsely; an exact test needs each cell's line, and doublets, called again from its genotype, which we could not do with the released data. The 75 treated samples without enough cells, mostly the most toxic conditions, are not used in the comparisons of predictions. We re-evaluated Rhaister, BioPert and the Molina--Zhang transfer model; State, whose released model was trained on both wells of the replicated drugs, was not re-evaluated, so how far its published scores are inflated is not shown here.

\section{Methods}

\textbf{Data and responses.} Single-cell counts of Tahoe-100M were taken from the public release\cite{zhang2026}. Each cell was assigned to one of two halves of its line and well by the lowest bit of the CRC32 hash of its barcode, and the mean of $\log(1+10^4\,c/n)$ over the cells of each half was computed per gene ($c$, a gene's count; $n$, the cell's total count). Lines: the 48 of 50 lines with a median of at least 40 cells per treated well. Samples: the 1,241 of 1,316 treated samples in which each of these lines has at least 10 cells in each half; most of the other 75 are the most toxic conditions. Genes: the 6,134 genes whose control expression, averaged over lines, plates and halves, is at least 0.1. A line's baseline expression is its mean over the control wells of all plates and both halves; a line's genes are those with baseline $\geq$ 0.2, and the common genes those whose median baseline over lines is $\geq$ 0.2. A line's response in a well is the mean of its two half-responses. References: (i) the line's mean over the cells of the plate's two control wells, each half against the same half of the controls (control-well reference, as in the published analyses); (ii) the line's mean over the plate's other treated wells, excluding the well itself, each half against the same half of those wells (treated-well reference; used throughout unless stated otherwise); (iii) the restricted reference, as (ii) but only over treated wells without detectable viability effect (below). Under the control-well reference the plate's control wells, with their own well effect, are subtracted from the target and from the same-well prediction alike (Results); a split version subtracts one control well from the target line and the other from the remaining lines, and the reverse, and averages the two; a version that subtracts the same single control well from every line has the same noise as the split version and keeps the shared term. The same two versions were computed for the treated-well reference, with the plate's other treated wells divided into two halves (alternate sample numbers) in place of the two control wells. Correlations for a line use the line's genes; variance components and signatures use the common genes, centred over genes. Well positions are inferred from sample numbers, assuming the samples of a plate are numbered row by row from A1 to H12; on the 13 plates used, the two control wells are the plate's last two samples, which places them at H11 and H12.

\textbf{Replicate wells and the label check.} Two wells are replicates if their drug and dose agree after salt names are removed (one trailing parenthesised qualifier and trailing salt words: citrate, hydrochloride, dihydrochloride, maleate, mesylate, sulfate, HCl, sodium; stereoisomers are kept apart); no pair involving plate 13 is used. Pairs are grouped as one plate; plates 6 and 14 at the same position; and different plates and positions, which include the six pairs between plates 6 and 14 of the three compounds present in two salt forms on each plate. Adagrasib at 0.05\,µM, present on every plate (two wells on plate 7), is reported apart. For each well, two readouts were compared, by Pearson correlation, with every treated well of the partner's plate except the partner and the well itself (plate 13 excluded): the line-averaged response under the treated-well reference (common genes, centred) and the viability readout, each line's log2 odds of its share of the well's cells (all 50 lines, with 0.5 added) minus the same for the plate's pooled control cells, centred over lines. The percentile of a pair in one direction is the share of these wells whose similarity is below the partner's. A pair is concordant if, for either readout, both directions reach the 90th percentile; the readouts are computed under the treated-well reference and are not recomputed for the other references. Decoys: in each of three classes (one plate, also split into adjacent wells and wells at least four positions apart; different plates other than 6 and 14; plates 6 and 14 at different positions), 5,000 random pairs of treated wells of different compounds were checked with the same rule; the number expected to pass by chance in a group if every label were wrong is the decoy rate times the group's number of pairs, and $p$ the one-sided binomial probability of at least the observed number. Variants: concordance by viability alone, by expression alone, and cross-fitted (expression signatures from one half of the cells, responses scored from the other half, in both directions). The viability spread of a well is the standard deviation over lines of the viability readout; a pair has no detectable viability effect if both wells lie below 0.289, the 90th percentile of the spread of one control well against the other (13 plates, both directions).

\textbf{Heat maps (Figs.~\ref{fig1} and \ref{fig3}).} Profiles are $\log(1+10^4c/N)$ of counts summed over a line's cells in a half or a whole well, on the protein-coding genes (as for BioPert, below); a well enters for a line if it has at least 50 cells and 35,000 counts and each half at least 10 cells. Genes: the 7,208 whose median over lines of the line's mean control profile is at least 0.2. Drug-treated profiles are taken against the line's mean over the plate's other qualifying treated wells (whole wells); control pairs need no reference. The 48 lines were split once at random into two sets of 24. For each pair of wells, the whole-well difference was averaged over the first set; the 150 genes with the highest and the 150 with the lowest mean were kept and ordered by it, and only lines of the second set are shown. A row of a two-well panel is half 1 of the first well minus half 2 of the second, a row of a one-well panel half 1 minus half 2 of the first well, so both compare half a well with half a well. A row's score is its Pearson correlation with the first set's mean over the kept genes; a pair is summarized by the median over its rows, and the examples are the pairs at the median of their group (the lower middle for an even number); at different positions, the median is taken among the concordant pairs whose lines' responses agree between the two wells at least at the group's median, so that the example's drug response reproduces. Checks on whole-well differences: thinning the deeper well's counts of each line binomially to the other well's total changes the control-pair scores by at most 0.011; regressing each line's difference on its mean response in the two wells, which removes a proportional difference in drug strength, changes the median scores from 0.93 to 0.92 (concordant pairs at different positions) and from 0.85 to 0.84 (plates 6 and 14). The lineup of Fig.~\ref{fig2} uses the similarities of the label check above.

\textbf{Same-well and replicate-well predictions.} For each pair of wells, each line measured in both wells and each of the two wells as the target's well, the same-well prediction is the mean response of the other lines in the target's well and the replicate-well prediction the mean of the other lines in the other well; neither includes the target line. The advantage is the difference of the two correlations. A pair is summarized by the medians over its lines and a group by the median over pairs; the advantage of a group is the mean over its records, a record being one line of one pair averaged over the two wells as the target's well (Supplementary Table~\ref{tabS2}). The compound-weighted advantage is the mean over compounds of each compound's mean record; signs over compounds are tested with a two-sided binomial test. Intervals are percentiles of 1,000 bootstrap resamples (2,000 for the well's share) of compounds (descriptive when a group has few compounds), or of compounds and plates independently (a record weighted by the product of its compound's weight and its plates' weights, a one-plate pair counting its plate once; a plate present in every record of a group is not resampled; resamples with zero total weight are dropped); we also report the range when each plate is left out in turn. For control pairs the response is the difference between the two control wells, and the replicate-well prediction is the mean difference of the other lines on each other plate, with correlations averaged over plates. Robustness of control pairs: profiles from counts summed over the line's cells in a well, $\log(1+10^6\,c/N)$ ($N$, the summed total), and the same after binomial thinning of the deeper well's summed counts to the depth of the other (per line and plate); the 13 $\times$ 13 table of medians over each plate's lines when predicted by each plate's pair, whose mean diagonal is compared with 100,000 random assignments of pairs to plates; and a bootstrap over plates (5,000 resamples) of the mean per-plate difference between the diagonal and the row's other entries.

\textbf{Permutation.} For each pair, every line's responses, the target's included, are swapped between the two wells with probability 1/2 and the statistic is recomputed; the statistic is the mean over pairs of the pair-mean advantage, computed exactly from Gram matrices of the centred responses over the genes expressed in all lines of the pair, so its observed value can differ slightly from the advantage, which uses each line's genes; 1,000 permutations; one-sided $p=(1+k)/1{,}001$, where $k$ permuted values reach the observed one; the tables give the largest permuted value.

\textbf{Variance components.} For a line in wells $a$ and $b$ with halves $A$ and $B$: $W$, the mean of the two wells' within-well half products, estimates the reproducible part of one well (drug effect and well effect); $Q$, the mean of the four cross-well half products, the drug effect; $S_\mathrm{same}$ and $S_\mathrm{other}$, the mean inner products of the line's response with the other lines' responses in the same well and in the other well. The drug share is $\Sigma Q/\Sigma W$, the well effect common to the lines $\Sigma(S_\mathrm{same}-S_\mathrm{other})/\Sigma W$, and the remainder the line-specific well effect; sums run over records, so records weigh by the size of their responses; the compound-weighted share averages the shares of compounds. Sampling noise is independent between halves and does not enter $W$ or $Q$. A difference in drug strength between the wells alone gives within-well products $W_a$, $W_b$ and a cross-well product $Q=\sqrt{W_aW_b}$, and same-well and cross-well line products related in the same way; replacing the arithmetic means of the two wells by these geometric means (negative within-well products set to zero; for line products of opposite sign the smaller is used) gives shares that a proportional difference cannot produce (Supplementary Table~\ref{tabS3}); computing the geometric means from sums over a pair's lines instead of per record changes them by at most 0.01. These are moment estimates of the variance components of the model $y_{lwh}=\mu_{ld}+u_w+v_{lw}+\varepsilon_{lwh}$ for the response of line $l$ in well $w$ (drug--dose $d$) and half $h$, each term a vector over the common genes: $\mu_{ld}$, the line's response to the drug, shared by every well of the condition; $u_w$, a well effect common to the lines of the well; $v_{lw}$, a well effect specific to the line; and $\varepsilon_{lwh}$, sampling noise. The four kinds of term are assumed uncorrelated with each other over genes ($\mu_{ld}$ may be correlated between lines), $u_w$ and $v_{lw}$ to have mean zero and to be independent between the two wells of a pair, $v_{lw}$ to be independent between lines, and the noise to be independent between halves, wells and lines. Then the expected product of two halves of one well is $\|\mu_{ld}\|^2+\mathrm{E}\|u_w\|^2+\mathrm{E}\|v_{lw}\|^2$ ($W$), that of halves of two wells $\|\mu_{ld}\|^2$ ($Q$), and the product of a line's response with the other lines' mean exceeds that from the other well by $\mathrm{E}\|u_w\|^2$ ($S_\mathrm{same}-S_\mathrm{other}$); the drug share, the common and the line-specific well share follow. The response is taken against a reference, the line's mean over the plate's other treated wells; a part of the reference that the two halves of a well share and the replicate well does not, such as the plate's average drug effect when the replicates lie on different plates, is counted with the well, and with the restricted reference, which leaves out most of that average, the well's share in concordant pairs at different positions is 0.38 instead of 0.35. A part of the well effect that both wells share, such as one tied to a shared position, is counted with $\mu_{ld}$; an interaction of plate and drug, which differs between the two wells, is counted with the well, and so would contamination between the lines of a well be, with the common well effect. Intervals come from resampling compounds and plates (above), not from distributional assumptions. A likelihood-based mixed model would add such assumptions but, like any model, cannot separate well from drug where a condition was measured in one well. Where $W$ is small, as without detectable viability effect, the ratio is unstable and its intervals reach beyond 1. In simulations with well effects drawn from combinations of real control-pair differences and noise from control-well half differences, the mean bias of the drug share was within 0.04 when pairs were not selected on the data themselves. The reported groups other than all pairs of plates 6 and 14 are selected: concordance keeps pairs whose wells agree, which raises $Q$ and biases the well's share in concordant pairs low; strong responses are selected on response length, which is related to $W$, so the share in strongly responding pairs is descriptive.

\textbf{Ceiling for a prediction from other wells.} Let a line's response in well $b$ be $y_b=s+e_b$ and in its replicate $y_a=s+e_a$, with $s$ the part that the two wells share and $e_a$, $e_b$ the parts they do not share (well effects and sampling noise), uncorrelated with $s$ and with each other, over genes. For a prediction $p$ uncorrelated with $e_b$, one that uses nothing from well $b$ or from wells that share part of $e_b$, such as neighbouring wells or the plate's control wells, $\mathrm{corr}(p,y_b)=\mathrm{corr}(p,s)\sqrt{\mathrm{var}(s)/\mathrm{var}(y_b)}\le\sqrt{\mathrm{var}(s)/\mathrm{var}(y_b)}$. If the two wells are equally noisy, $\mathrm{var}(y_a)=\mathrm{var}(y_b)$, the correlation between the two wells is $r_{ab}=\mathrm{var}(s)/\mathrm{var}(y_b)$ and the bound is $\sqrt{r_{ab}}$. In general the bound is $\sqrt{r_{ab}}\,(\mathrm{var}(y_a)/\mathrm{var}(y_b))^{1/4}$, below $\sqrt{r_{ab}}$ when the scored well is the noisier of the two and above it otherwise. The same holds for two plates in place of two wells. Whatever the two wells share counts as reproducible here, so replicates that share a position give a bound for the drug's effect together with what is tied to the position. A prediction that borrows from well $b$ is correlated with $e_b$ and can exceed the bound.

\textbf{Alternative explanations.} Dose direction: for a pair of a compound at one dose, each line's dose direction is its mean response in the compound's wells at the higher of the two other doses minus that at the lower (the pair's wells and plate 13 excluded); every response of the line in either well is replaced by its least-squares residual on this direction before scoring, and the share of the line's difference between the two wells explained by it is reported. Spatial structure: for each position of plates 6 and 14, the difference between the two plates' responses averaged over lines (common genes); correlations between positions at Chebyshev distance 1, 2--3 and $\geq$ 4 (pairs of positions with different compounds); the median at distance 1 minus that at $\geq$ 4 was compared with 1,000 shuffles of the profiles over the positions. Adagrasib: for each of its 14 wells and each line, the same-well prediction minus the mean over its other wells of the prediction from that well; sign test over wells. Contamination: a line's foreign genes are protein-coding genes whose expression in its pooled control cells (counts per million of summed counts) is below 5\% of the median over the other lines, with that median at least 10 and at least 25 other lines at 10 or more; the contamination fraction $f$ of a line in a well is the share of its counts on these genes divided by the same share in the pooled counts of the other lines in the well; a line with fewer than 10 foreign genes takes the median $f$ of the other lines in the well. The fraction is an upper bound, because a foreign gene's own low expression is counted as contamination. Decontaminated response, to first order: $(y-f\,m)/(1-f)$, with $m$ the mean response of the other lines in the well, applied to every line before scoring, and again with every fraction doubled. Quality regression: on 0.05\,µM wells of compounds without replicates, per line and gene, slopes on the line's mean counts per cell and cell number in the well and on the well's mitochondrial fraction and genes per cell; row and column effects removed by median polish. Ambient and composition directions: median $R^2$ of a line's difference between the two wells of a pair regressed on the well's average baseline expression minus the line's own (control-well reference), on the baseline weighted by the difference in line composition between the wells (treated-well reference), and on the mean difference of all lines (each alone).

\textbf{Model evaluation.} Test lines were the five held out by State (C32, HOP-62, HepG2/C3A, Hs 766T, PANC-1). Drug--doses were keyed by salt-stripped names. For each drug--dose measured in more than one well, only one well entered training (plate 6 for plates 6/14, otherwise the lowest plate number; and, in a second run, the other well) and the others never did; plate 13 was not used. In the few-shot setting each test line also contributed a random 30\% of its single-well conditions. The perturbation mean is the mean response of the training lines in the training well; the additive model fits line and drug--dose terms by 10 rounds of alternating means. Scores are correlations of the prediction with the test line's response in the training well and in the other well, for concordant pairs, with the treated-well reference unless stated otherwise. Rhaister was run with its public code and inputs on its five held-out splits\cite{rhaister2026}, with each replicated drug--dose kept in its inputs only on the training plate, so the other plate never entered training, and the truth taken from the training plate or from another plate; its ceiling is the same metric computed between the two plates' observed differential expression of the same records.

\textbf{Models with a measured-response input.} BioPert\cite{kaufmann2026}: profiles were rebuilt from raw counts as in its code (counts summed over 19,020 protein-coding genes per line and well; at least 50 cells and 35,000 counts; $\log(1+10^4\,c/n)$), with responses relative to the line's mean over the plate's other treated wells or, as in BioPert, to the plate's control wells. The drug input is A549's response in one well of the drug--dose and the line input the line's mean control profile. The network (principal components of both inputs and a multilayer perceptron with layers of 1,024, 128 and 1,024 units) and its training schedule were ported to PyTorch with the hyperparameters selected for Tahoe-100M; on the published compound split the port reproduced the reported test correlation (0.79). Every compound with a replicate pair was held out for testing, together with its stereoisomers and other salts in the data (three compounds). For each pair of wells, in both directions and for every line measured in both wells, the predictions from A549's response in the target's well and in the other well were scored against the target's response in its own well. With the treated-well reference BioPert was trained with three seeds (medians within 0.02 of each other), with the control-well reference with one. The fingerprint model (ECFP6, 2,048 bits) kept its published hyperparameters, which were selected with the control-well reference. Molina--Zhang transfer model\cite{molina2026}: State's five lines; the source is the mean response of the other four lines in a well (at least three present), mapped to the target line by a 30-dimensional basis of the training sources and ridge regression, both fitted on compounds without replicate wells and their stereoisomers or salts; the penalty was chosen on 20\% of these compounds from $10^{-4}$ to $10^{3}$ times the mean feature variance and lay inside this range for all five lines. The prediction from the source in one well was scored against the target's response in the same well and in the other well, in both directions of each pair. Pairs are those of plates 6 and 14 and at different positions with concordant labels, and those without detectable viability effect, concordant or not; plate 13 was not used.

\textbf{Calibration of the cell-cycle test.} As in the Tahoe-100M analysis\cite{zhang2026}: for each line, the difference between the log2 odds ratios of the G1 and the G2/M proportions in the treated against the reference well; mean over lines; two-sided $p$ from 10,000 sign flips. In the original design the treated well is compared with the plate's two pooled control wells; we applied it to treated wells without detectable viability effect whose drugs' annotated targets and mechanisms (broad and fine) contain none of a list of terms for the cell cycle, DNA damage, mitosis, transcription, chromatin and major signalling pathways, among drugs with an annotation (599 wells kept; 452 excluded for a viability effect, 171 for their mechanism or a missing annotation). As a drug-free comparison, one control well was compared with the other.

\textbf{Plate 13.} Partner percentiles and viability spreads as above for the 45 plate-13 pairs. Response length is the Euclidean norm of the line-averaged response (common genes, centred); a well responds strongly if its length is $\geq$ 3 (for recognition, thresholds 2.5 and 4 give the same picture). Length ratio: in every group the first well is the reference well (for plate 13, the well on the other plate; for plates 6 and 14, the plate-6 well; otherwise the well on the lower-numbered plate, or with the lower sample number on one plate); a pair enters if the first well's length is $\geq$ 3, and the ratio is the partner's length over the first well's. Counts are given with the number of compounds and are not tested, because pairs of one compound are not independent. Adagrasib responses in plate-13 wells were compared with its wells on other plates in the most responsive lines.

\textbf{Other data.} Emerald Bay\cite{emeraldbay}: one 96-well plate of the same platform (51--52 lines per well); the four DMSO wells (first-round barcodes 54, 85, 86 and 87; their positions were not checked); per cell $\log(1+10^4\,c/n)$, mean per line and well, lines with at least 30 cells in each of the four wells (48 lines); line genes with mean $\geq$ 0.2 over the four wells; for each of the six pairs the difference of the lower-numbered barcode minus the higher, predicted by the other lines of the same pair and by those of the complementary pair. OP3\cite{szalata2024}: in each donor, 16 wells each of DMSO, belinostat and dabrafenib; same-well against another-well prediction of a cell type by the other five types. sci-Plex3\cite{srivatsan2020}, 24-hour data: 2,256 pairs of replicate wells, all at different positions; correlation between two halves of one well against halves of two replicate wells with equal cell numbers, relative to the plate's other treated wells.

\textbf{Software.} Python 3.11 with NumPy, SciPy, PyArrow and scikit-learn; Rhaister with its own environment (Polars, PyTorch); BioPert ported to PyTorch, with RDKit for fingerprints; figures with Matplotlib.

\backmatter
\bmhead{Supplementary information}
Supplementary Fig.~\ref{figS1} and Supplementary Tables~\ref{tabCtrl}--\ref{tabS5} follow the references.
\bmhead{Data availability}
Tahoe-100M is available from Hugging Face (\url{https://huggingface.co/datasets/tahoebio/Tahoe-100M}), Emerald Bay from Hugging Face (\url{https://huggingface.co/datasets/tahoebio/EmeraldBay}), Rhaister's inputs from Hugging Face (\url{https://huggingface.co/datasets/tahoebio/tahoe-de-rhaister}), OP3 from the Open Problems repository and sci-Plex3 from GEO (GSE139944). No new data were generated. The replicate pairs of Tahoe-100M with the outcome of the label check (concordance, decoy rates, plate 13) are in the code repository (files \texttt{data/labels84.json}, \texttt{data/ver99.json} and \texttt{data/p13strong92.json}), for evaluations that need verified replicate wells.
\bmhead{Code availability}
Analysis code and the result files behind every figure and table are available at \url{https://github.com/xuanruihuang/well-is-the-unit} under the MIT licence; the figures can be redrawn from the included result files, and the analysis rerun from the public data with the included scripts; the supplementary tables were compiled by hand from the result files named in the repository's README.
\bmhead{Ethics}
This study analysed only publicly released data.

\bmhead{Author contributions}
X.H. conceived the study, performed the analyses and wrote the manuscript.

\bmhead{Competing interests}
The author declares no competing interests.

\clearpage
\section*{Supplementary information}
\FloatBarrier
\renewcommand{\tablename}{Supplementary Table}
\setcounter{table}{0}
\renewcommand{\figurename}{Supplementary Fig.}
\setcounter{figure}{0}
\begin{figure}[!htbp]
\centering
\makebox[\textwidth][c]{\includegraphics{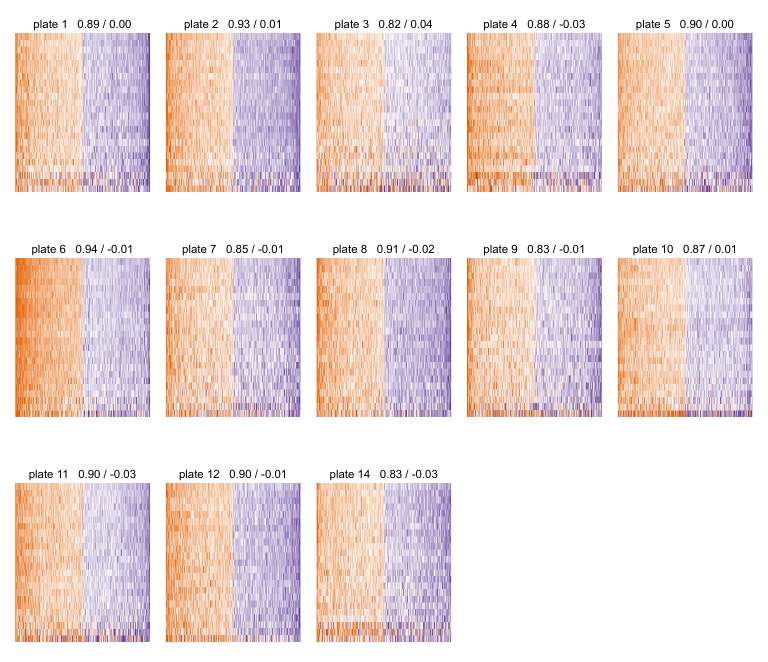}}
\caption{\textbf{All 13 control-well pairs.} As Fig.~\ref{fig1}b, for each plate (plate 13 not used); each panel on its own colour scale. Above each panel, the median correlation of a shown line with the choosing lines' mean between the two wells and, after the slash, between the two halves of control well 1.}\label{figS1}
\end{figure}
\begin{table}[!htbp]
\sbox{\tbox}{\begin{minipage}{183mm}
\caption{\textbf{Untreated control-well pairs.} Response: a line's difference between a plate's two DMSO wells (H11 minus H12). Same pair: correlation, over the line's genes, with the mean difference of the other lines in the same pair; other plates: with that of the other lines in the control pairs of the other plates, averaged over plates; medians over line $\times$ plate records. Advantage: mean over records of the difference. Permutation: the mean diagonal of the 13 $\times$ 13 table in which each plate's lines are predicted by each plate's pair, against 100,000 random assignments of pairs to plates (the smallest possible $p$ is $10^{-5}$). Plate bootstrap: mean over plates of the plate's diagonal minus its other entries, with a 95\% interval from 5,000 resamples of plates. Profiles: mean over cells of $\log(1+10^4c/n)$ (main text); counts summed over the line's cells in a well, $\log(1+10^6c/N)$; the same after thinning the deeper well's counts to the depth of the other. First-order decontamination: Methods. Plate 13 (not used in the text; its labels are unreliable, and its control wells, inferred from sample numbers, sit at B11 and B12, so the low prediction from other plates may reflect position or plate 13's other problems): its own pair against the pairs of the other plates. Emerald Bay: one later plate of the same platform; its four DMSO wells split into two pairs in three ways; each pair's lines are predicted by the pair's other lines and by those of the other pair of the split; medians over 288 line $\times$ pair records. The splits share wells and are not independent.}\label{tabCtrl}
\centering\fontsize{7}{8.4}\selectfont\setlength{\tabcolsep}{4pt}
\begin{tabular}{@{}>{\raggedright\arraybackslash}p{58mm}rrrrrr@{}}\toprule
Pairs and profile & Same pair & Other plates & Advantage & Plates higher & Permutation $p$ & Plate bootstrap \\
\midrule
Tahoe-100M, 13 plates, 624 records & & & & & & \\
\quad mean over cells of log-expression & 0.70 & 0.47 & 0.20 & 13 of 13 & $10^{-5}$ & 0.22 (0.19--0.25) \\
\quad summed counts, log counts per million & 0.68 & 0.46 & 0.19 & 13 of 13 & $10^{-5}$ & 0.20 (0.16--0.25) \\
\quad summed counts, matched depth & 0.66 & 0.45 & 0.18 & 13 of 13 & $10^{-5}$ & 0.19 (0.15--0.24) \\
\quad mean over cells, first-order decontamination & 0.67 & & & & & \\
Plate 13, its own pair against other plates' pairs (auxiliary) & 0.77 & 0.10 & & & & \\
\quad summed counts; matched depth & 0.72; 0.66 & 0.10; 0.09 & & & & \\
\midrule
 & Same pair & Other pair & & Splits, both pairs higher & & \\
\midrule
Emerald Bay, 3 splits of 4 DMSO wells into two pairs, 48 lines & 0.65 & 0.24 & & 3 of 3 & & \\
\bottomrule\end{tabular}
\end{minipage}}
\typeout{TBL tabCtrl \the\wd\tbox\space\the\dimexpr\ht\tbox+\dp\tbox\relax\space\the\textheight}
\makebox[\textwidth][c]{\usebox{\tbox}}
\end{table}

\begin{table}[!htbp]
\sbox{\tbox}{\begin{minipage}{183mm}
\caption{\textbf{Replicate wells and the label check.} Replicates: wells with the same drug (salt names removed) and dose. Concordant: the partner well ranks in the top 10\% of the partner plate's treated wells in both directions, by the line-averaged response or by the viability readout. Chance: pairs expected to pass if every label of the group were wrong, from 5,000 decoy pairs of two different compounds per class (pass rates: one plate 12\%, of which adjacent wells 25\%; different plates 10\%; plates 6 and 14 at different positions 11\%); for plates 6 and 14 the decoys differ in position, while a shared position also helps a pair pass. No viability effect: both wells' viability spread below the 90th percentile of the spread between two control wells. Partner percentile: rank of the partner by similarity of the line-averaged response among the partner plate's treated wells (50, chance), median over pairs of the mean of both directions. Strong pairs matched: matched / strongly responding pairs, where a pair responds strongly if either well has a response length $\geq$ 3 and is matched if each well ranks the other among the top 10\% of that plate's treated wells by expression; descriptive (pairs of one compound are not independent; plates 6 and 14 share positions). Length ratio: where the first well responds strongly, the partner's response length over the first well's, median (number of pairs); the first well is the one on the other plate for plate 13, the plate-6 well for plates 6 and 14, otherwise the well on the lower-numbered plate or with the lower sample number; descriptive. A compound can appear in more than one row. Plate-13 pairs are not used in any analysis.}\label{tabS1}
\centering\fontsize{7}{8.4}\selectfont\setlength{\tabcolsep}{3pt}
\begin{tabular}{@{}>{\raggedright\arraybackslash}p{50mm}rrrrrrrr@{}}\toprule
Replicate wells & Pairs & Compounds & Concordant pairs & Chance & No viability effect & Partner & Strong pairs & Length ratio \\
 & & & (compounds) & & pairs (compounds) & percentile & matched & (pairs) \\
\midrule
Different plates and positions & 65 & 20 & 18 (9) & 6.4 & 24 (15) & 81 & 12 / 24 & \\
\quad of which different salt names (different plates) & 56 & 18 & 12 (8) & & & 78 & 6 / 17 & 0.79 (14) \\
\quad of which same names, plates 1--3 and 7--12 & 9 & 3 & 6 (2) & & & 96 & 6 / 7 & 0.77 (6) \\
Two wells on one plate & 24 & 7 & 6 (4) & 3.0 & 11 (6) & 48 & 0 / 5 & 0.97 (3) \\
Plates 6 and 14 (same position) & 90 & 87 & 80 (77) & 9.5 & 65 (63) & 100 & 17 / 18 & 0.71 (18) \\
Adagrasib 0.05\,µM, all pairs of its wells & 91 & 1 & 60 & & 78 & 98 & -- & \\
Plate 13 against other plates (not used) & 45 & 18 & & & & 44 & 1 / 10 & 0.44 (6) \\
\bottomrule\end{tabular}
\end{minipage}}
\typeout{TBL tabS1 \the\wd\tbox\space\the\dimexpr\ht\tbox+\dp\tbox\relax\space\the\textheight}
\makebox[\textwidth][c]{\usebox{\tbox}}
\end{table}

\begin{table}[!htbp]
\sbox{\tbox}{\begin{minipage}{183mm}
\caption{\textbf{Same-well against replicate-well prediction.} Treated-well reference unless stated. Same / replicate: correlation of the target line's response with the other lines' mean in its own well and in the replicate well; median over lines within a pair, then median over pairs. Advantage: mean difference over records, with 95\% intervals from 1,000 bootstrap resamples of compounds and of compounds and plates; range when each plate is left out in turn. Compound-weighted: mean over compounds of each compound's mean, with the number of compounds with a positive mean in parentheses (two-sided sign test; for the control-well rows of plates 6 and 14, of the split version). Permutation: the largest of 1,000 values of the mean advantage over pairs after swapping every line's responses between the two wells of a pair at random ($p=1/1{,}001$ in every group); its statistic uses the genes expressed in all lines of a pair, and its observed value lies within 0.02 of the advantage. Restricted: the reference restricted to the plate's treated wells without detectable viability effect. Concordance variants: by viability alone, by expression alone, and cross-fitted (signatures from one half of the cells, scores from the other; two directions, both given). Control-well reference: pooled control wells; one control well subtracted from every line (same noise as the split version, shared term kept); split, one control well subtracted from the target line and the other from the remaining lines, averaged over both assignments. Treated wells in halves: the same two versions with the plate's other treated wells divided into two halves in place of the two control wells.}\label{tabS2}
\centering\fontsize{7}{8.4}\selectfont\setlength{\tabcolsep}{1.8pt}
\begin{tabular}{@{}>{\raggedright\arraybackslash}p{34mm}rrrrrrrrr@{}}\toprule
Pairs & Pairs & Same & Replicate & Advantage & Compounds; & Leave one & Compound- & Permutation & Restricted \\
 & (compounds) & & & & and plates & plate out & weighted & max & reference \\
\midrule
Different positions, concordant & 18 (9) & 0.68 & 0.38 & 0.32 & 0.26--0.48; 0.18--0.51 & 0.28--0.35 & 0.38 (9/9) & 0.02 & 0.34 \\
\quad concordant by viability alone & 9 (6) & & & 0.29 & & & 0.35 (6/6) & & \\
\quad concordant by expression alone & 16 (7) & & & 0.29 & & & 0.33 (7/7) & & \\
\quad cross-fitted & 18; 19 (9; 10) & & & 0.30; 0.30 & & & 0.35; 0.36 (all) & & \\
\quad control wells: pooled; one; split & 18 (9) & & & 0.30; 0.30; 0.26 & & & & & \\
\quad treated wells in halves: one; split & 18 (9) & & & 0.32; 0.31 & & & 0.39; 0.36 (9/9) & & \\
Different positions, no viability effect & 24 (15) & 0.73 & 0.08 & 0.60 & 0.54--0.67; 0.45--0.77 & 0.56--0.65 & 0.60 (15/15) & 0.02 & 0.62 \\
\quad control wells: pooled; one; split & 24 (15) & & & 0.47; 0.45; 0.37 & & & & & \\
One plate, concordant & 6 (4) & 0.69 & 0.17 & 0.51 & 0.42--0.75; 0.29--0.84 & 0.40--0.56 & 0.56 (4/4) & 0.05 & 0.53 \\
One plate, no viability effect & 11 (6) & 0.70 & 0.04 & 0.62 & 0.48--0.73; 0.43--0.81 & 0.59--0.64 & 0.63 (6/6) & 0.04 & 0.64 \\
Plates 6 and 14 & 90 (87) & 0.70 & 0.48 & 0.24 & 0.22--0.27; -- & -- & 0.24 (87/87) & 0.004 & 0.25 \\
\quad concordant pairs & 80 (77) & 0.71 & 0.50 & 0.21 & 0.19--0.23; -- & -- & 0.21 (77/77) & & \\
\quad concordant by viability alone & 17 (17) & & & 0.22 & & & 0.22 (17/17) & & \\
\quad concordant, cross-fitted & 80 (77) & & & 0.19; 0.19 & & & 0.19 (77/77) & & \\
\quad strongly responding & 18 (18) & & & 0.20 & & & & & \\
\quad concordant, control wells: pooled; one; split & 80 (77) & & & 0.18; 0.23; 0.08 & & & 0.08 (70/77) & & \\
\quad concordant, treated wells in halves: one; split & 80 (77) & & & 0.22; 0.19 & & & 0.19 (77/77) & & \\
Adagrasib 0.05\,µM, 14 wells & & & & 0.37 & & & 14/14 wells & & \\
\bottomrule\end{tabular}
\end{minipage}}
\typeout{TBL tabS2 \the\wd\tbox\space\the\dimexpr\ht\tbox+\dp\tbox\relax\space\the\textheight}
\makebox[\textwidth][c]{\usebox{\tbox}}
\end{table}

\begin{table}[!htbp]
\sbox{\tbox}{\begin{minipage}{183mm}
\caption{\textbf{Sensitivity to the concordance threshold, and pairs by compound.} Treated-well reference; same-well and replicate-well predictions as in Supplementary Table~\ref{tabS2} (medians over lines for each pair, then medians over pairs). Top: pairs concordant with the partner among the top 5\%, 10\% (main text) or 20\% of the partner plate's treated wells, by expression or by viability, in both directions for the same readout; difference, mean over pairs of the pair's same-well minus replicate-well prediction (the main text's advantage is a mean over lines and pairs). A looser threshold also admits more pairs with wrong labels. Bottom: the concordant and no-viability-effect groups of Supplementary Table~\ref{tabS2} by compound (salt names removed); medians over the compound's pairs. Plates 6 and 14 (87 compounds) are not listed.}\label{tabS6}
\centering\fontsize{7}{8.4}\selectfont\setlength{\tabcolsep}{4pt}
\begin{tabular}{@{}llrrrr@{}}\toprule
Threshold & Group & Pairs (compounds) & Same & Replicate & Difference \\
\midrule
Top 5\% & Different positions & 14 (6) & 0.68 & 0.43 & 0.27 \\
Top 5\% & One plate & 3 (2) & 0.70 & 0.36 & 0.47 \\
Top 5\% & Plates 6 and 14 & 77 (75) & 0.71 & 0.50 & 0.21 \\
Top 10\% & Different positions & 18 (9) & 0.68 & 0.38 & 0.33 \\
Top 10\% & One plate & 6 (4) & 0.69 & 0.17 & 0.51 \\
Top 10\% & Plates 6 and 14 & 80 (77) & 0.71 & 0.50 & 0.22 \\
Top 20\% & Different positions & 36 (16) & 0.68 & 0.24 & 0.44 \\
Top 20\% & One plate & 11 (7) & 0.70 & 0.14 & 0.56 \\
Top 20\% & Plates 6 and 14 & 85 (82) & 0.70 & 0.49 & 0.23 \\
\bottomrule\end{tabular}

\vspace{3mm}
\begin{tabular}{@{}llrrr@{}}\toprule
Group & Compound & Pairs & Same & Replicate \\
\midrule
Concordant, different positions & berbamine & 1 & 0.74 & 0.38 \\
 & binimetinib & 3 & 0.68 & 0.29 \\
 & crizotinib & 1 & 0.73 & -0.13 \\
 & docetaxel & 1 & 0.64 & 0.22 \\
 & irinotecan & 1 & 0.80 & 0.44 \\
 & neratinib & 1 & 0.64 & 0.19 \\
 & tirabrutinib & 1 & 0.72 & 0.52 \\
 & tofacitinib & 1 & 0.68 & 0.37 \\
 & trametinib & 8 & 0.67 & 0.46 \\
\addlinespace[2pt]
Concordant, one plate & cytarabine & 1 & 0.67 & 0.16 \\
 & gallic acid & 1 & 0.71 & -0.14 \\
 & omeprazole & 1 & 0.70 & 0.18 \\
 & tofacitinib & 3 & 0.68 & 0.36 \\
\addlinespace[2pt]
No viability effect, different positions & afatinib & 1 & 0.76 & 0.24 \\
 & almonertinib & 2 & 0.73 & 0.10 \\
 & belumosudil & 2 & 0.72 & 0.15 \\
 & berbamine & 4 & 0.74 & 0.01 \\
 & bestatin & 1 & 0.69 & 0.03 \\
 & canagliflozin & 1 & 0.76 & 0.32 \\
 & crizotinib & 2 & 0.74 & -0.14 \\
 & dapagliflozin & 1 & 0.72 & -0.24 \\
 & docetaxel & 1 & 0.69 & 0.10 \\
 & fluvoxamine & 1 & 0.72 & 0.09 \\
 & gallic acid & 2 & 0.72 & 0.03 \\
 & irinotecan & 1 & 0.73 & -0.06 \\
 & neratinib & 2 & 0.64 & 0.12 \\
 & pyridoxine & 1 & 0.74 & 0.26 \\
 & tofacitinib & 2 & 0.69 & 0.28 \\
\addlinespace[2pt]
No viability effect, one plate & cytarabine & 1 & 0.67 & 0.16 \\
 & gallic acid & 2 & 0.71 & -0.10 \\
 & gemcitabine & 1 & 0.70 & -0.04 \\
 & omeprazole & 2 & 0.71 & -0.01 \\
 & pyridoxine & 2 & 0.70 & -0.01 \\
 & tofacitinib & 3 & 0.68 & 0.36 \\
\bottomrule\end{tabular}
\end{minipage}}
\typeout{TBL tabS6 \the\wd\tbox\space\the\dimexpr\ht\tbox+\dp\tbox\relax\space\the\textheight}
\makebox[\textwidth][c]{\usebox{\tbox}}
\end{table}

\begin{table}[!htbp]
\sbox{\tbox}{\begin{minipage}{183mm}
\caption{\textbf{Alternative explanations.} Quality and layout: advantage of the same-well prediction on plates 6 and 14 (90 pairs) with the control-well reference (which includes the shared control-well term, Supplementary Table~\ref{tabS2}), before and after regression on the line's mean counts per cell and cell number in the well and the well's mitochondrial fraction and genes per cell (slopes per line and gene fitted on 0.05\,µM wells of compounds without replicates), and after removal of row and column effects of the plate (median polish). Product and dose: concordant pairs at different positions split by name; advantage before and after removing each line's dose direction (difference between its responses to the compound's other two doses) from its responses in both wells, and the median share of a line's difference between the two wells explained by that direction. Contamination: estimated fraction of a line's counts from other lines (an upper bound for the 3 lines with at least 10 foreign genes, the median of the well for the others; median and 90th percentile over records), and the advantage before and after first-order removal, also with every fraction doubled; control pairs: same-pair prediction. Relation between the estimated fraction and the advantage over records (Spearman correlation; advantage below and above the median fraction): concordant pairs at different positions 0.43 (0.25, 0.40); plates 6 and 14, concordant 0.02 (0.21, 0.21); control-well pairs, with the same-pair prediction, $-0.29$. Ambient and composition: median $R^2$ of a line's difference between the two wells of a pair regressed on the ambient direction (the well's average baseline expression minus the line's own; control-well reference), on the baseline weighted by the difference in line composition between the wells (treated-well reference), and on the mean difference of all lines between the wells (the well effect itself, for comparison).}\label{tabS7}
\centering\fontsize{7}{8.4}\selectfont\setlength{\tabcolsep}{4pt}
\begin{tabular}{@{}>{\raggedright\arraybackslash}p{70mm}rrrr@{}}\toprule
Quality and layout (plates 6 and 14, control-well reference) & Advantage & & & \\
\midrule
None & 0.19 & & & \\
Regression on quality measures & 0.19 & & & \\
Removal of row and column effects & 0.18 & & & \\
\midrule
Product and dose (treated-well reference) & Pairs (compounds) & Advantage & After dose direction & $R^2$ of dose direction \\
\midrule
Different positions, concordant & 18 (9) & 0.32 & 0.36 & 0.02 \\
\quad same names & 6 (2) & 0.32 & & \\
\quad different salt names & 12 (8) & 0.32 & & \\
One plate, concordant & 6 (4) & 0.51 & 0.57 & 0.01 \\
Plates 6 and 14, concordant & 79 (76) & 0.21 & 0.21 & 0.02 \\
\midrule
Contamination (treated-well reference) & Fraction & Advantage & After removal & Fraction doubled \\
\midrule
Different positions, concordant & 0.060 / 0.076 & 0.32 & 0.31 & 0.30 \\
Different positions, no viability effect & 0.064 / 0.086 & 0.60 & 0.58 & 0.56 \\
One plate, concordant & 0.073 / 0.089 & 0.51 & 0.49 & 0.47 \\
One plate, no viability effect & 0.073 / 0.088 & 0.62 & 0.59 & 0.56 \\
Plates 6 and 14, concordant & 0.060 / 0.075 & 0.21 & 0.20 & 0.19 \\
Control pairs (same-pair prediction) & 0.054 / 0.069 & 0.70 & 0.67 & 0.65 \\
\midrule
Ambient direction (control-well reference) & Ambient & Mean difference of lines & & \\
\midrule
Different positions, all & 0.042 & 0.47 & & \\
One plate, all & 0.024 & 0.51 & & \\
Plates 6 and 14 & 0.020 & 0.35 & & \\
\midrule
Composition direction (treated-well reference) & Composition & Mean difference of lines & & \\
\midrule
Different positions, concordant & 0.002 & 0.52 & & \\
Different positions, no viability effect & 0.002 & 0.55 & & \\
One plate, no viability effect & 0.002 & 0.55 & & \\
Plates 6 and 14 & 0.003 & 0.37 & & \\
Control pairs (no reference) & 0.002 & 0.52 & & \\
\bottomrule\end{tabular}
\end{minipage}}
\typeout{TBL tabS7 \the\wd\tbox\space\the\dimexpr\ht\tbox+\dp\tbox\relax\space\the\textheight}
\makebox[\textwidth][c]{\usebox{\tbox}}
\end{table}

\begin{table}[!htbp]
\sbox{\tbox}{\begin{minipage}{183mm}
\caption{\textbf{What two halves of a well share: the well's share.} Treated-well reference unless stated. Pooled: $1-\Sigma Q/\Sigma W$ over records, which weights records by the size of their responses; 95\% interval from 2,000 bootstrap resamples of compounds and plates (not truncated; on plates 6 and 14 both plates are in every record and only compounds are resampled). Compound-weighted: mean over compounds of each compound's share. Common part: the well effect common to the lines, $\Sigma(S_\mathrm{same}-S_\mathrm{other})/\Sigma W$. Without strength difference: with the geometric instead of the arithmetic mean of the two wells, which removes any contribution of a proportional difference in drug strength between the wells (Methods). Where $W$ is small (no viability effect, and the six concordant pairs on one plate) the intervals reach beyond 1 and the values are descriptive. Strongly responding: response length $\geq$ 3 in either well.}\label{tabS3}
\centering\fontsize{7}{8.4}\selectfont\setlength{\tabcolsep}{3.5pt}
\begin{tabular}{@{}>{\raggedright\arraybackslash}p{62mm}rrrrrr@{}}\toprule
Pairs (pairs, compounds) & \multicolumn{3}{c}{Well's share} & Common part & \multicolumn{2}{c}{Without strength difference} \\ \cmidrule(lr){2-4}\cmidrule(lr){6-7}
 & Pooled & Interval & Compound-weighted & & Well's share & Common part \\
\midrule
Different positions, concordant (18, 9) & 0.35 & 0.22--0.58 & 0.51 & 0.22 & 0.32 & 0.21 \\
\quad without trametinib (10, 8) & 0.44 & & 0.54 & 0.30 & & \\
Different positions, no viability effect (24, 15) & 0.79 & 0.59--1.08 & 0.84 & 0.55 & 0.78 & 0.55 \\
One plate, concordant (6, 4) & 0.56 & 0.30--1.17 & 0.70 & 0.40 & 0.54 & 0.39 \\
One plate, no viability effect (11, 6) & 0.77 & 0.47--1.10 & 0.84 & 0.51 & 0.76 & 0.52 \\
Plates 6 and 14 (90, 87) & 0.30 & 0.18--0.49 & 0.45 & 0.16 & 0.25 & 0.14 \\
\quad concordant pairs (80, 77) & 0.24 & 0.15--0.41 & 0.40 & 0.13 & 0.20 & 0.11 \\
\quad pairs with a strongly responding well (18, 18) & 0.21 & & & 0.10 & 0.14 & 0.07 \\
Control-well reference & & & & & & \\
\quad different positions, concordant & 0.41 & & & 0.25 & 0.38 & 0.23 \\
\quad different positions, no viability effect & 0.72 & & & 0.43 & 0.71 & 0.43 \\
\quad one plate, concordant & 0.21 & & & 0.22 & 0.20 & 0.22 \\
\quad one plate, no viability effect & 0.34 & & & 0.28 & 0.33 & 0.28 \\
\bottomrule\end{tabular}
\end{minipage}}
\typeout{TBL tabS3 \the\wd\tbox\space\the\dimexpr\ht\tbox+\dp\tbox\relax\space\the\textheight}
\makebox[\textwidth][c]{\usebox{\tbox}}
\end{table}

\begin{table}[!htbp]
\sbox{\tbox}{\begin{minipage}{183mm}
\caption{\textbf{Model evaluation against the well that supplied training data or inputs and against a replicate well.} Top: held-out cell lines. State's five test lines; for each replicated drug--dose (salt names removed) only one well entered training; plate 13 not used; few-shot. Scores: median Pearson correlation over test line $\times$ drug--dose records, against the well whose other lines entered training and against the replicate well; concordant pairs only. Training well: plate 6 for plates 6/14 and otherwise the lowest plate number, or, in the second run, the other well. Agreement of the test lines' own responses between the two wells over the same records (first run, treated wells): 0.54 on plates 6 and 14, 0.66 at different positions. Reference: treated wells (treated-well reference) or control wells (control-well reference; on plates 6 and 14 it shares the control wells between training well and target). For one-plate pairs the first well is the one with the lower sample number. The number of compounds differs between the runs because a record needs the test line measured in both wells and the drug--dose kept out of its training conditions in that direction. Rhaister: its own metric (correlation of predicted and observed differential expression, each plate against its own control wells), on the drugs of plates 6 and 14, against the training plate and the other plate; ceiling, the same metric between the two plates' observed values. Bottom: models whose input is a measured response, on compounds held out from training (all compounds with a replicate well, and their stereoisomers and salts). BioPert: input, A549's response in the target's well (same well) or in the replicate well; 47 target lines; seed 42 (three seeds with the treated-well reference differ by at most 0.02); ECFP model, BioPert's fingerprint model (ECFP6), which uses no well and whose hyperparameters were selected with the control-well reference. Molina--Zhang transfer model (low-rank ridge): the mean response of the other four of State's five lines in a well, scored against the target in the same well and in the replicate well. Both directions of each pair; own replicate, the target line's response in the replicate well (it differs between the two models because their lines, records and genes differ); medians over line $\times$ pair-direction records; genes expressed in the line.}\label{tabS4}
\centering\fontsize{7}{8.4}\selectfont\setlength{\tabcolsep}{3pt}
\begin{tabular}{@{}lllrrrr@{}}\toprule
 & & & \multicolumn{2}{c}{Additive model} & \multicolumn{2}{c}{Perturbation mean} \\ \cmidrule(lr){4-5}\cmidrule(lr){6-7}
Pairs & Reference & Training well & Training well & Replicate well & Training well & Replicate well \\
\midrule
Plates 6 and 14 (77 compounds) & treated wells & plate 6 & 0.74 & 0.51 & 0.74 & 0.51 \\
 & treated wells & plate 14 & 0.72 & 0.51 & 0.72 & 0.50 \\
 & control wells & plate 6 & 0.82 & 0.60 & 0.73 & 0.48 \\
Different positions (6 compounds) & treated wells & lower plate & 0.70 & 0.35 & 0.70 & 0.36 \\
 & treated wells & higher plate (9 compounds) & 0.68 & 0.39 & 0.68 & 0.39 \\
 & control wells & lower plate & 0.75 & 0.42 & 0.71 & 0.37 \\
One plate (3 compounds) & treated wells & first well & 0.71 & 0.25 & 0.71 & 0.25 \\
 & treated wells & second well (4 compounds) & 0.72 & 0.10 & 0.72 & 0.10 \\
 & control wells & first well & 0.73 & 0.47 & 0.68 & 0.43 \\
\bottomrule\end{tabular}

\vspace{3mm}
\begin{tabular}{@{}lrrr@{}}\toprule
Rhaister, plates 6 and 14 (1,588 records) & Training plate & Other plate & Ceiling (the two plates' observations) \\
\midrule
Few-shot & 0.93 & 0.65 & 0.67 \\
Zero-shot & 0.65 & 0.44 & 0.67 \\
\bottomrule\end{tabular}

\vspace{3mm}\setlength{\tabcolsep}{2.2pt}
\begin{tabular}{@{}>{\raggedright\arraybackslash}p{58mm}lrrrrrrr@{}}\toprule
 & & \multicolumn{4}{c}{BioPert} & \multicolumn{3}{c}{Transfer} \\ \cmidrule(lr){3-6}\cmidrule(lr){7-9}
Pairs (pairs, compounds) & Reference & Same & Replicate & Own & ECFP & Same & Replicate & Own \\
 & & well & well & replicate & model & well & well & replicate \\
\midrule
Plates 6 and 14, concordant (80, 77) & treated wells & 0.63 & 0.43 & 0.46 & 0.07 & 0.71 & 0.49 & 0.54 \\
Different positions, concordant (18, 9) & treated wells & 0.73 & 0.51 & 0.51 & 0.04 & 0.88 & 0.68 & 0.67 \\
Different positions, no viability effect (24, 15) & treated wells & 0.63 & 0.03 & 0.09 & 0.01 & 0.69 & 0.08 & 0.14 \\
One plate, no viability effect (11, 6) & treated wells & 0.60 & 0.03 & 0.02 & $-$0.02 & 0.66 & $-$0.01 & 0.09 \\
Plates 6 and 14, concordant & control wells & 0.75 & 0.58 & 0.55 & 0.33 & 0.79 & 0.62 & 0.63 \\
Different positions, concordant & control wells & 0.84 & 0.48 & 0.50 & 0.39 & 0.89 & 0.63 & 0.65 \\
Different positions, no viability effect & control wells & 0.78 & 0.19 & 0.18 & 0.41 & 0.82 & 0.21 & 0.21 \\
One plate, no viability effect & control wells & 0.76 & 0.48 & 0.50 & 0.38 & 0.80 & 0.43 & 0.51 \\
\bottomrule\end{tabular}
\end{minipage}}
\typeout{TBL tabS4 \the\wd\tbox\space\the\dimexpr\ht\tbox+\dp\tbox\relax\space\the\textheight}
\makebox[\textwidth][c]{\usebox{\tbox}}
\end{table}

\begin{table}[!htbp]
\sbox{\tbox}{\begin{minipage}{183mm}
\caption{\textbf{Reproducibility ceilings and the cell-cycle test.} Top: agreement of a line's responses between two random halves of one well, Spearman--Brown corrected (mean of the two wells), and between the two whole replicate wells; median over lines within a pair, then median over pairs; treated-well reference; groups as in Supplementary Table~\ref{tabS2}. Bottom: the sign-flip test of the Tahoe-100M cell-cycle analysis (mean over lines of the difference between the log2 odds ratios of the G1 and the G2/M proportions; two-sided $p$ from 10,000 sign flips over lines). Control-well pairs: one control well of a plate against the other (13 plates, plate 13 not used; significant plates have signs in both directions). Original design: each treated well against the plate's two pooled control wells, for wells without detectable viability effect whose drugs' annotated targets and mechanisms contain no term for the cell cycle, DNA, mitosis, transcription, chromatin or major signalling pathways (a keyword filter; Methods); the wells of a plate share its control wells and are not independent.}\label{tabCeil}
\centering\fontsize{7}{8.4}\selectfont\setlength{\tabcolsep}{4pt}
\begin{tabular}{@{}>{\raggedright\arraybackslash}p{70mm}rrr@{}}\toprule
Pairs & Pairs & Two halves of one well & Two wells \\
\midrule
Concordant, different positions & 18 & 0.94 & 0.57 \\
Concordant, one plate & 6 & 0.84 & 0.35 \\
No viability effect, different positions & 24 & 0.80 & 0.10 \\
No viability effect, one plate & 11 & 0.84 & 0.06 \\
Plates 6 and 14 & 90 & 0.81 & 0.48 \\
\bottomrule\end{tabular}

\vspace{3mm}
\begin{tabular}{@{}>{\raggedright\arraybackslash}p{70mm}rrr@{}}\toprule
Cell-cycle test & Tests & $p<0.05$ & $p<0.01$ \\
\midrule
Control-well pairs, one well against the other & 13 plates & 5 & 4 \\
Original design, wells of drugs unrelated to the cell cycle (238 compounds) & 599 wells & 68\% & 56\% \\
\quad range over the 13 plates & & 48--93\% & \\
\bottomrule\end{tabular}
\end{minipage}}
\typeout{TBL tabCeil \the\wd\tbox\space\the\dimexpr\ht\tbox+\dp\tbox\relax\space\the\textheight}
\makebox[\textwidth][c]{\usebox{\tbox}}
\end{table}

\begin{table}[!htbp]
\sbox{\tbox}{\begin{minipage}{183mm}
\caption{\textbf{Other pooled screens.} OP3: six immune cell types of one donor per well; within each donor, 16 wells each of DMSO, belinostat and dabrafenib; a cell type's response predicted by the mean of the other five types in the same well or in another well of the same donor and compound; replicate agreement, the cell type's own response in two wells; medians over records. sci-Plex3: one cell line per well, 24-hour data, 2,256 pairs of replicate wells at different positions; correlation between two halves of one well against halves of two replicate wells with equal cell numbers (median), difference with 95\% interval over compounds, and the share of pairs in which halves of one well agree more.}\label{tabS5}
\centering\fontsize{7}{8.4}\selectfont\setlength{\tabcolsep}{4pt}
\begin{tabular}{@{}llrrrr@{}}\toprule
OP3 & Records & Replicate agreement & Same well & Another well & \\
\midrule
DMSO & 1,080 & $-$0.05 & 0.27 & 0.00 & \\
Dabrafenib & 1,080 & 0.13 & 0.18 & 0.05 & \\
Belinostat & 775 & 0.78 & 0.53 & 0.50 & \\
\midrule
sci-Plex3 & Reference & Halves of one well & Halves of two wells & Difference & One well higher \\
\midrule
 & control wells & 0.036 & 0.017 & 0.019 (0.017--0.021) & 74\% \\
 & treated wells & 0.075 & 0.041 & 0.031 (0.028--0.033) & 84\% \\
\bottomrule\end{tabular}
\end{minipage}}
\typeout{TBL tabS5 \the\wd\tbox\space\the\dimexpr\ht\tbox+\dp\tbox\relax\space\the\textheight}
\makebox[\textwidth][c]{\usebox{\tbox}}
\end{table}


\begin{thebibliography}{10}
\expandafter\ifx\csname url\endcsname\relax
  \def\url#1{\burl{#1}}\fi
\expandafter\ifx\csname urlprefix\endcsname\relax\def\urlprefix{URL }\fi
\providecommand{\bibinfo}[2]{#2}
\providecommand{\eprint}[2][]{\url{#2}}
\providecommand{\doi}[1]{\url{https://doi.org/#1}}
\bibcommenthead

\bibitem{bunne2024}
\bibinfo{author}{Bunne, C.}, \bibinfo{author}{Roohani, Y.},
  \bibinfo{author}{Rosen, Y.} \emph{et~al.}
\newblock \bibinfo{title}{{How to build the virtual cell with artificial
  intelligence: priorities and opportunities}}.
\newblock \emph{\bibinfo{journal}{Cell}} \textbf{\bibinfo{volume}{187}},
  \bibinfo{pages}{7045--7063} (\bibinfo{year}{2024}).

\bibitem{roohani2025}
\bibinfo{author}{Roohani, Y.~H.}, \bibinfo{author}{Hua, T.~J.},
  \bibinfo{author}{Tung, P.-Y.} \emph{et~al.}
\newblock \bibinfo{title}{{Virtual Cell Challenge: toward a Turing test for the
  virtual cell}}.
\newblock \emph{\bibinfo{journal}{Cell}} \textbf{\bibinfo{volume}{188}},
  \bibinfo{pages}{3370--3374} (\bibinfo{year}{2025}).

\bibitem{mcfarland2020}
\bibinfo{author}{McFarland, J.~M.}, \bibinfo{author}{Paolella, B.~R.}
  \emph{et~al.}
\newblock \bibinfo{title}{{Multiplexed single-cell transcriptional response
  profiling to define cancer vulnerabilities and therapeutic mechanism of
  action}}.
\newblock \emph{\bibinfo{journal}{Nat. Commun.}} \textbf{\bibinfo{volume}{11}},
  \bibinfo{pages}{4296} (\bibinfo{year}{2020}).

\bibitem{zhang2026}
\bibinfo{author}{Zhang, J.}, \bibinfo{author}{Ubas, A.~A.} \emph{et~al.}
\newblock \bibinfo{title}{{Tahoe-100M: mapping drug-induced molecular
  phenotypes at single-cell resolution}}.
\newblock \emph{\bibinfo{journal}{Cell}} \textbf{\bibinfo{volume}{189}},
  \bibinfo{pages}{5945--5961} (\bibinfo{year}{2026}).

\bibitem{shen2026}
\bibinfo{author}{Shen, H.}
\newblock \bibinfo{title}{{Basal expression predicts context-specific drug
  responses only where they are reproducible: a leakage-safe 47-line LOCO
  benchmark on Tahoe-100M}}.
\newblock \emph{\bibinfo{journal}{Research Square}}  (\bibinfo{year}{2026}).
\newblock \bibinfo{note}{\url{https://doi.org/10.21203/rs.3.rs-10846736/v1}}.

\bibitem{hurlbert1984}
\bibinfo{author}{Hurlbert, S.~H.}
\newblock \bibinfo{title}{{Pseudoreplication and the design of ecological field
  experiments}}.
\newblock \emph{\bibinfo{journal}{Ecol. Monogr.}}
  \textbf{\bibinfo{volume}{54}}, \bibinfo{pages}{187--211}
  (\bibinfo{year}{1984}).

\bibitem{zimmerman2021}
\bibinfo{author}{Zimmerman, K.~D.}, \bibinfo{author}{Espeland, M.~A.} \&
  \bibinfo{author}{Langefeld, C.~D.}
\newblock \bibinfo{title}{{A practical solution to pseudoreplication bias in
  single-cell studies}}.
\newblock \emph{\bibinfo{journal}{Nat. Commun.}} \textbf{\bibinfo{volume}{12}},
  \bibinfo{pages}{738} (\bibinfo{year}{2021}).

\bibitem{squair2021}
\bibinfo{author}{Squair, J.~W.}, \bibinfo{author}{Gautier, M.},
  \bibinfo{author}{Kathe, C.} \emph{et~al.}
\newblock \bibinfo{title}{{Confronting false discoveries in single-cell
  differential expression}}.
\newblock \emph{\bibinfo{journal}{Nat. Commun.}} \textbf{\bibinfo{volume}{12}},
  \bibinfo{pages}{5692} (\bibinfo{year}{2021}).

\bibitem{adduri2025}
\bibinfo{author}{Adduri, A.~K.} \emph{et~al.}
\newblock \bibinfo{title}{{Predicting cellular responses to perturbation across
  diverse contexts with State}}.
\newblock \emph{\bibinfo{journal}{bioRxiv}}  (\bibinfo{year}{2025}).
\newblock \bibinfo{note}{\url{https://doi.org/10.1101/2025.06.26.661135}}.

\bibitem{gandhi2025}
\bibinfo{author}{Gandhi, S.}, \bibinfo{author}{Javadi, F.},
  \bibinfo{author}{Svensson, V.} \emph{et~al.}
\newblock \bibinfo{title}{{Tahoe-x1: scaling perturbation-trained single-cell
  foundation models to 3 billion parameters}}.
\newblock \emph{\bibinfo{journal}{bioRxiv}}  (\bibinfo{year}{2025}).
\newblock \bibinfo{note}{\url{https://doi.org/10.1101/2025.10.23.683759}}.

\bibitem{rhaister2026}
\bibinfo{author}{{Tahoe Therapeutics}}.
\newblock \bibinfo{title}{{Rhaister: model card, dataset card and holdout split
  definitions}}.
\newblock \bibinfo{howpublished}{Hugging Face, tahoebio/Rhaister and
  tahoebio/tahoe-de-rhaister} (\bibinfo{year}{2026}).
\newblock \bibinfo{note}{Accessed 20 October 2026}.

\bibitem{kaufmann2026}
\bibinfo{author}{Kaufmann, L.}, \bibinfo{author}{Le~Breton, L.} \emph{et~al.}
\newblock \bibinfo{title}{{A biological-response compound representation allows
  chemical perturbation prediction across cell lines}}.
\newblock \emph{\bibinfo{journal}{bioRxiv}}  (\bibinfo{year}{2026}).
\newblock \bibinfo{note}{\url{https://doi.org/10.64898/2026.09.28.755146}}.

\bibitem{molina2026}
\bibinfo{author}{Molina, A.} \& \bibinfo{author}{Zhang, X.}
\newblock \bibinfo{title}{{Perturbation response decomposition enables
  biologically aligned generalization to unseen perturbations and cellular
  contexts}}.
\newblock \emph{\bibinfo{journal}{bioRxiv}}  (\bibinfo{year}{2026}).
\newblock \bibinfo{note}{\url{https://doi.org/10.64898/2026.07.24.740459}}.

\bibitem{celleval}
\bibinfo{author}{{Arc Institute}}.
\newblock \bibinfo{title}{{cell-eval: evaluation suite for perturbation
  response prediction}}.
\newblock \bibinfo{howpublished}{GitHub, ArcInstitute/cell-eval}
  (\bibinfo{year}{2025}).
\newblock \bibinfo{note}{Accessed 20 October 2026}.

\bibitem{wang2026}
\bibinfo{author}{Wang, X.}, \bibinfo{author}{Kuipers, J.},
  \bibinfo{author}{Hugi, F.}, \bibinfo{author}{Platt, R.~J.} \&
  \bibinfo{author}{Beerenwinkel, N.}
\newblock \bibinfo{title}{{Reliable single-cell perturbations explain and
  improve model performance}}.
\newblock \emph{\bibinfo{journal}{bioRxiv}}  (\bibinfo{year}{2026}).
\newblock \bibinfo{note}{\url{https://doi.org/10.64898/2026.08.11.744177}}.

\bibitem{mazoure2017}
\bibinfo{author}{Mazoure, B.}, \bibinfo{author}{Nadon, R.} \&
  \bibinfo{author}{Makarenkov, V.}
\newblock \bibinfo{title}{{Identification and correction of spatial bias are
  essential for obtaining quality data in high-throughput screening
  technologies}}.
\newblock \emph{\bibinfo{journal}{Sci. Rep.}} \textbf{\bibinfo{volume}{7}},
  \bibinfo{pages}{11921} (\bibinfo{year}{2017}).

\bibitem{kalinin2025}
\bibinfo{author}{Kalinin, A.~A.}, \bibinfo{author}{Arevalo, J.},
  \bibinfo{author}{Serrano, E.} \emph{et~al.}
\newblock \bibinfo{title}{{A versatile information retrieval framework for
  evaluating profile strength and similarity}}.
\newblock \emph{\bibinfo{journal}{Nat. Commun.}} \textbf{\bibinfo{volume}{16}},
  \bibinfo{pages}{5181} (\bibinfo{year}{2025}).

\bibitem{emeraldbay}
\bibinfo{author}{{Tahoe Therapeutics}}.
\newblock \bibinfo{title}{{EmeraldBay}}.
\newblock \bibinfo{howpublished}{Hugging Face, tahoebio/EmeraldBay}
  (\bibinfo{year}{2026}).
\newblock \bibinfo{note}{Accessed 29 October 2026}.

\bibitem{spearman1910}
\bibinfo{author}{Spearman, C.}
\newblock \bibinfo{title}{{Correlation calculated from faulty data}}.
\newblock \emph{\bibinfo{journal}{Br. J. Psychol.}}
  \textbf{\bibinfo{volume}{3}}, \bibinfo{pages}{271--295}
  (\bibinfo{year}{1910}).

\bibitem{brown1910}
\bibinfo{author}{Brown, W.}
\newblock \bibinfo{title}{{Some experimental results in the correlation of
  mental abilities}}.
\newblock \emph{\bibinfo{journal}{Br. J. Psychol.}}
  \textbf{\bibinfo{volume}{3}}, \bibinfo{pages}{296--322}
  (\bibinfo{year}{1910}).

\bibitem{szalata2024}
\bibinfo{author}{Sza{\l}ata, A.}, \bibinfo{author}{Benz, A.},
  \bibinfo{author}{Cannoodt, R.} \emph{et~al.}
\newblock \bibinfo{title}{{A benchmark for prediction of transcriptomic
  responses to chemical perturbations across cell types}}.
\newblock \emph{\bibinfo{journal}{Adv. Neural Inf. Process. Syst. (Datasets and
  Benchmarks Track)}}  (\bibinfo{year}{2024}).

\bibitem{srivatsan2020}
\bibinfo{author}{Srivatsan, S.~R.}, \bibinfo{author}{McFaline-Figueroa, J.~L.},
  \bibinfo{author}{Ramani, V.} \emph{et~al.}
\newblock \bibinfo{title}{{Massively multiplex chemical transcriptomics at
  single-cell resolution}}.
\newblock \emph{\bibinfo{journal}{Science}} \textbf{\bibinfo{volume}{367}},
  \bibinfo{pages}{45--51} (\bibinfo{year}{2020}).

\bibitem{huang2026clone}
\bibinfo{author}{Huang, X.}
\newblock \bibinfo{title}{{Shared clones make random cell halves look
  reproducible in pooled screens}}.
\newblock \emph{\bibinfo{journal}{arXiv}}  (\bibinfo{year}{2026}).
\newblock \bibinfo{note}{\url{https://arxiv.org/abs/2610.32962}}.

\end{thebibliography}
\end{document}